\documentclass[%
 reprint,
 amsmath,amssymb,
 aps,
]{revtex4-2}

\usepackage{amsmath}
\usepackage{braket}
\usepackage{graphicx}% Include figure files
\usepackage{dcolumn}% Align table columns on decimal point
\usepackage{bm}% bold math
\begin{document}

\preprint{APS/123-QED}

\title{Tuning and Mapping of Two-Tone Optomechanical Instability up to the Physical Limits}% Force line breaks with \\

% \affil[1]{Université Grenoble Alpes, Institut Néel, France}
% \affil[2]{Aalto University, Finland}

% \author{Marco Dicosta}
\date{July 2026}

\author{Marco Dicosta$^1$}
\author{Alexandre Delattre$^1$}
%\author{Mika Sillanpää $^{2}$}
\author{Louise Banniard$^2$}
\author{Laure Mercier de L\'epinay$^2$}
\author{Andrew Fefferman$^1$}
\author{Eddy Collin$^{1,\dagger}$}
% \email{eddy.collin@neel.cnrs.fr}
\affiliation{$^1$ Institut N\'eel UGA - CNRS, \\ 25 rue des Martyrs, 38042 Grenoble, France}
\affiliation{$^2$ Department of Applied Physics, Aalto University, Aalto, Finland \\}

\date{\today}% It is always \today, today,

\begin{abstract}
We report on the compete theory describing the two-tone instability proper to optomechanical back-action-evading experiments. We address all possible mistunings (including power imbalance of the two pumps), as well as the device imperfections (finite mechanical quality factor, and sideband resolution). 
We illustrate the importance of power mistuning and sideband resolution, the latter being particularly relevant in finding the back-action-evading "sweet spot".
The model is compared to experiments performed up to an extremely high cooperativity, of almost a 100.
This has been made possible thanks to a fine-tuning procedure, which is presented.  
The agreement theory-experiment is very good, and
the limiting factor is unraveled: cavity frequency noise.
Besides, we discover at very high drive powers new self-driven states, which all share the characteristic feature of presenting a mechanical comb spectrum;
as if the synchronization of the responses to the two drives (which occurs at lower powers) was destroyed by nonlinear effects.
We believe that our work is extremely relevant to the  implementation of back-action-evading experiments, as well as other related two-tone schemes.
\end{abstract}

\maketitle

%\tableofcontents

\section{Introduction}

Optomechanics has proven over the years to be an extremely sensitive detection tool, finding widespread application across a diverse range of fields. The first direct gravitational wave detection has been 
achieved with a gigantic interferometer \cite{popoXXX}.   At the other end of the length-scale spectrum, micro- and nano-optomechanical platforms can perform sensing down to almost the molecular level, for instance using water with applications in nanofluidics \cite{lalaXXX}.  
%%%
Shifting down the photon energies to the GigaHertz range, the more cryogenic-compatible microwave version of optomechanics has emerged about two decades ago \cite{regal}. This has lead to unique quantum-oriented experiments, like the entanglement of two distinct mechanical resonators \cite{entangle}.    

The basic principles of cavity optomechanics are well known \cite{AKM}.
Applying a pump tone detuned downwards from the cavity resonance frequency 
by exactly the mechanical resonance frequency (a "red-detuned" pump), one performs cooling (via back-action damping) \cite{teufel}. 
 The reverse process is a tone detuned above the cavity resonance ("blue-detuned"), producing 
amplification (via anti-damping) \cite{zhou}. These phenomena increase with injected pump power.
With the blue-detuned scheme, when the total damping on the mechanics reaches zero the system starts to self-oscillate \cite{DylanPRR}: this is the (single-tone) parametric instability. 
Damping and anti-damping come along with a frequency shift, the optical spring effect \cite{AKM}.

 Continuous motion detection through a single pump tone applied at the cavity frequency is intrinsically bound by quantum mechanics (the back-action induced "quantum limit") \cite{AKM}. 
But more elaborate schemes are possible, as pointed out originally in Refs. \cite{Braginsky,Thorne}: with a blue-detuned {\it and} a red-detuned pump applied simultaneously, one can overcome this limit by routing all added noise towards the unmeasured quadrature, while continuously monitoring the other one. 
%%%CORRR!!!
The complete optomechanics theory of this so-called back-action-evading technique (for {\it ideal tuning conditions}) was further developed in Ref. \cite{JacobsNJP}, and first tested experimentally in Ref.  \cite{schwabBAE}.
Obviously, the signal-to-noise improves as one increases the optical power.

But the  experimental implementation of this principle requires extreme care in the tuning of all parameters related to the two pump fields. Eventually, the system will even start to self-oscillate: this is the {\it two-tone instability} we are interested in here.
It has been first studied in Ref. \cite{KippenbergPRX}, both experimentally and theoretically. This earlier work pointed out the relevance of pump frequency mistunings, and that the tuning conditions become stricter as input power is increased. The Authors realized that an experimental limit emerges at the highest powers, but did not qualify it any further.

We present here the full theory underlying this two-tone instability, and compare it to experiments.
All possible mistunings are taken into account (including the power imbalance of the two pumps which had not been addressed so far), as well as the device imperfections (finite quality factor, sideband resolution parameter). {\it All of these imperfections} are experimentally relevant, if one is to stabilize a proper back-action evading measurement at high powers. 
We present a fine-tuning technique (that builds on the theoretical model) which enables us to work up to extremely large injected power, essentially up to the physical limit of the system. We demonstrate that this ultimate limit is the $1/f$-type intrinsic noise in the cavity resonance properties, which is in the community believed to emerge from Two-Level-System (TLS) defects \cite{burnett}.  
%%%%%%%%%%%%%%%%%%%%%
At these large powers, a zoo of new self-oscillating states appears, which are all characterized by the emergence of {\it a comb in the mechanical spectrum}. 
At low injected powers, the two pumps always end up generating a single self-oscillating peak, as if "synchronized". Understanding how this can be destroyed when the power is too large (certainly because of nonlinearities) would bring new insights in the optically-driven synchronization properties, as studied e.g. in Ref. \cite{synchro}.
Our work is  also relevant to other similar 
two-tone schemes  \cite{lehnert,mikastuff}.

\section{Theory}

%The presented theory captures the full features on the shape and construction of the parametric instabilities, showing the importance for correct tuning of the input fields and fundamentally demonstrating an asymmetry due to the sideband resolution. 
%%% this is something for a conclusion/introduction. Not here.
In order to construct the full theoretical treatment of both the tuning ability and instabilities, we consider the Hamiltonian \cite{AKM}:
%\\
\begin{equation}
    \hat{\mathcal{H}}=\hat{\mathcal{H}}_0+\hat{\mathcal{H}}_{int}+\hat{\mathcal{H}}_{Drive}+\hat{\mathcal{H}}_{baths} \label{hamilton} , 
\end{equation}
%\\
%\\
where $\hat{\mathcal{H}}_0=\hbar \omega_c \hat{a}^\dag \hat{a}+\hbar \Omega_m \hat{b}^\dag \hat{b}$ is the initial Hamiltonian with cavity boson operators $\hat{a}$ and mechanics $\hat{b}$, $\hat{\mathcal{H}}_{int}=-\hbar g_0 \hat{a}^\dag \hat{a} \left(  \hat{b}^\dag + \hat{b}  \right)$ is the generic interaction Hamiltonian (coupling strength $g_0$) and $\hat{\mathcal{H}}_{baths}$ is the Hamiltonian for the baths.
%\\
$\hat{\mathcal{H}}_{Drive}$ consists of the drive fields for the red and blue pump $\hat{\mathcal{H}}_{Drive}= \hat{\mathcal{H}}_{red} + \hat{\mathcal{H}}_{blue}$, with:
\begin{equation}
    \hat{\mathcal{H}}_{red} = -i\hbar\sqrt{\kappa_{ext}}\,[\hat{a}^\dagger \alpha_R(t)-\hat{a}\alpha_R^*(t)] ,
\end{equation}
\begin{equation}
    \hat{\mathcal{H}}_{blue} = -i\hbar\sqrt{\kappa_{ext}}\,[\hat{a}^\dagger \alpha_B(t)-\hat{a}\alpha_B^*(t)] ,
\end{equation}
%\\
%\\
where $\alpha_R(t)=\frac{\alpha_0}{\sqrt{2}}(1-\epsilon)e^{-i\omega_Rt+\phi_R}$  and $\alpha_B(t)=\frac{\alpha_0}{\sqrt{2}}(1+\epsilon)e^{-i\omega_Bt+\phi_B}$ in which $\epsilon$ stands for the %ratio of 
input pumps amplitude mistuning, a positive $\epsilon$ indicates that the blue input field is tuned to a higher injected power than the red input field (and vice-versa). $\kappa_{ext}$ is the external coupling rate of the cavity and is defined via $\kappa_{tot}=\kappa_{ext}+\kappa_{in}$, where $\kappa_{tot}$ is the total damping rate %and linewidth 
of the cavity and $\kappa_{in}$ corresponds to the internal losses. %damping of the cavity. 
%\\
The positions of the pumps in frequency are defined as $\omega_{R,B}$  where $\omega_R=\omega_p-\omega_m$ defines the frequency position of the red pump and $\omega_B=\omega_p+\omega_m$  the frequency position of the blue pump.
$\omega_p$ and $\omega_m$ are the pump frequencies relative to the cavity and mechanical resonances. %detuning relative to the mechanical frequency, 
Both are defined via mistunings parameters with $\omega_p= \omega_c+\Delta$ where $\omega_c$ is the actual cavity resonance frequency and $\Delta$ is the cavity mistuning, and $\omega_m= \Omega_m + \delta$ where $\Omega_m$ is the resonance frequency of the mechanical element and $\delta$ is the mechanical mistuning.
%by construction positive for detunings for the red pump and negative for the blue pump, this ensures that as $\delta\rightarrow\infty$ the projected mechanics spacing grows at the same rate. 
The input field phases are recast into  
$\phi=\frac{\phi_B+\phi_R}{2}$ and $\delta\phi=\frac{\phi_B-\phi_R}{2}$; they play essentially no role here, as will be seen in the expressions below. % that other noise elements play more significant roles.
\\
\\
% Injecting the red and blue input fields into input drive field we obtain:
We can recast the input fields into a single {\it amplitude modulated}  field:
%\\
%\\
$\alpha_{in}(t)=\sqrt{2}\alpha_0e^{-i\phi}e^{-i\omega_pt}[\cos(\omega_mt+\delta\phi)-i\epsilon\sin(\omega_mt+\delta\phi)]$.   
%\\
%\\
%with the defined drive field, 
In order to solve the problem, we apply a double-rotating wave approximation around $\omega_m$ and $\omega_p$ plus a displacement operator, using $\hat{a}(t)=\tilde{a}(t)e^{-i\omega_pt}$ with $\tilde{a}(t)=\delta \tilde{a}(t) +\tilde{\alpha}(t)$ for optics, and $\hat{b}=\tilde{b}(t)e^{-i\omega_mt}$ for the mechanics as the ansatz.
The Hamiltonian becomes:
%\\
%\\
\begin{eqnarray}
    \tilde{\mathcal{H}}=-\hbar\Delta \, \delta\tilde{a}^\dagger\delta\tilde{a}-\hbar\delta \, \tilde{b}^\dagger\tilde{b}\nonumber\\
    -\hbar g_0(\tilde{\alpha}\delta\tilde
    {a}^\dagger+\tilde{\alpha}^*\delta\tilde{a})(\tilde{b}^\dagger e^{+i\omega_mt}+\tilde{b}e^{-i\omega_mt})\nonumber\\
    -\hbar\Delta(\tilde{\alpha}\delta\tilde{a}^\dagger+\tilde{\alpha}^*\delta \tilde{a})+i\hbar (\dot{\tilde{\alpha}}^*\delta\tilde{a}-\dot{\tilde{\alpha}}\delta\tilde{a}^\dagger)\nonumber\\
    -i\hbar\sqrt{\kappa_{ext}}[\delta\tilde{a}^\dagger\tilde{\alpha}_{in}-\delta\tilde{a}\tilde{\alpha}_{in}^*]+\tilde{\mathcal{H}}_{Baths} .
\end{eqnarray}
%\\
%\\
We have also introduced the drive field in the rotating frame $\alpha_{in}(t)=\tilde{\alpha}_{in}(t)e^{-i\omega_pt}$.
The Quantum Langevin Equations in the double rotating frame can be written in terms of the detunings in cavity and mechanical frequencies:
\begin{eqnarray}
        \delta\dot{\tilde{a}}=+i\Delta\delta\tilde{a}+ig_0\tilde{\alpha}(t)\left(\tilde{b}^\dagger e^{i\omega_mt}+\tilde{b}e^{-i\omega_mt}\right)\nonumber\\
        -\frac{\kappa_{tot}}{2}\delta\tilde{a}+\left[ \mbox{Bath terms} \right] ,
    \end{eqnarray}
%%%%%%%
\begin{eqnarray}
        \dot{\tilde{\alpha}}(t)=+i\Delta\, \tilde{\alpha}(t)-\frac{\kappa_{tot}}{2} \tilde{\alpha}(t)-\sqrt{\kappa_{ext}}\tilde{\alpha}_{in}(t) ,
    \end{eqnarray}
for the optical field, having split the operator component from the complex number expectation value,
\begin{eqnarray}
        \dot{\tilde{b}}=+i\delta\,\tilde{b}+ig_0 \left(\tilde{\alpha}(t)\delta\tilde{a}^\dagger+\tilde{\alpha}^*(t)\delta\tilde{a} \right)e^{+i\omega_mt}\nonumber\\-\frac{\Gamma_m}{2}\tilde{b}+\left[ \mbox{Bath terms} \right] ,
\end{eqnarray}
%\\
for the mechanics. 
%%%%   SHOULD DEFINE IT!
The mechanical damping rate is $\Gamma_m$, and the corresponding quality factor $Q_0=\Omega_m/\Gamma_m$.
%%%%
We first solve for $\tilde{\alpha}$, and will reinject the result into the $\delta \tilde{a}$ and $\tilde{b}$ equations above.
% Solving for $\Tilde{\alpha}$
We obtain the expression for the %input power:
intracavity field:
\begin{widetext}
\begin{equation}
    \tilde{\alpha}=-\sqrt{\kappa_{ext}}\sqrt{2} \,\alpha_0e^{-i\phi} \cdot \left[\frac{(i(\Delta-\epsilon\,\omega_m)-\frac{\kappa_{tot}}{2})\cos{(\omega_mt+\delta\phi)+(\epsilon \,\Delta-\omega_m+i\epsilon\frac{\kappa_{tot}}{2}})\sin{(\omega_mt+\delta\phi)}}{(\Delta^2-\omega_m^2-\frac{\kappa_{tot}^2}{4}) +i (\Delta\kappa_{tot})}\right] .
\end{equation}
\end{widetext}
%\noindent
This result is {\it very different} from the formula which can be found in the literature \cite{KippenbergPRX,JacobsNJP}. 
%in contemporary work which omits the finite $\epsilon$ and $\kappa_{tot}$ terms.
Indeed, we have introduced here the power mistuning parameter $\epsilon$, and we 
  maintain a $\kappa_{tot}$ inbuilt dependence, which allows to take into account the effect of a finite sideband resolution 
%%%% HERE DEFINE IT!!!
$S=\kappa_{tot}/\Omega_m$. % to be inbuilt into the equation. 
Thus, the expression given above is {\it the most generic} for the input field,  incorporating all possible mistunings ($\delta, \Delta$ and $\epsilon$) as well as %capturing the effect of the sideband resolution parameter. 
device imperfection (namely $S$). The finite mechanical $Q_0$ itself appears in Eq. (\ref{dynmat}) below. 
Injecting the cavity field operator into the $ \tilde{b} $ equation the {\it dynamical matrix} can be constructed:
%\\
%\\
\begin{widetext}
    \begin{equation}
\begin{bmatrix}
   \frac{\Gamma_m}{2}+\frac{\gamma(\omega)}{2}+i(-\omega-\delta)-i\lambda(\omega) & +\Sigma(\omega)\\
   +\Sigma^*(-\omega)&   \frac{\Gamma_m}{2}+\frac{\gamma(-\omega)}{2}-i(\omega-\delta)+i\lambda(-\omega)
\end{bmatrix}
\begin{bmatrix}
    \tilde{b}(\omega)\\ \tilde{b}^\dagger(\omega)
\end{bmatrix}
=
\begin{bmatrix}
   \mbox{Bath}\\\mbox{terms}
\end{bmatrix} . \label{dynmat}
\end{equation}
\end{widetext}
%
%\\
%\\
%\noindent
%Where $\Sigma(\omega)$ is the self energy of system.
%%% do you know what a self-energy is??
%%% I would not call that like this; I would simply say:
The functions $\gamma(\omega) , \lambda(\omega)$ and  $\Sigma(\omega)$ are given in Appendix \ref{theorystuff}.
Inverting the dynamical matrix provides the susceptibility $\chi_m(\omega)$ of the mechanics, 
%%% I would immediately ADD:
while diagonalizing it we can look for the zeroes of the real part which defines the instability limit.
We finally construct the spectra from Kubo's notation \cite{ClerkDevoretRevModPhys}:
\begin{equation}
\begin{split}
        S_{\tilde{b}}(\omega)= \int\frac{\braket{\tilde{b}(\omega)\tilde{b}^\dagger(-\omega')}}{2\pi}d\omega' ,\\
        S_{\tilde{b}^\dagger}(\omega)=\int\frac{\braket{\tilde{b}^\dagger(\omega)\tilde{b}(-\omega')}}{2\pi}d\omega' ,
\end{split}
\end{equation}
%\\
for the mechanics, and for 
the optical spectrum it is defined via:
\begin{equation}
\begin{split}
        S_{\delta \tilde{a}}(\omega)= \int\frac{\braket{\delta\tilde{a}(\omega)\delta\tilde{a}^\dagger(-\omega')}}{2\pi}d\omega' ,\\
        S_{\delta\tilde{a}^\dagger} (\omega)=\int\frac{\braket{\delta\tilde{a}^\dagger(\omega)\delta\tilde{a}(-\omega')}}{2\pi}d\omega' .
\end{split}
\end{equation}
%\\
%The spectrum is injected into the optics, 
%The terms relating to 
%%%%%%%% I rewrite things the right way...
 
Since the cavity is near its ground state and the phonon population is large $n_{ph} \gg 1$, we neglect here the quantum fluctuations and the cavity noise back-action (precisely the quantity of interest for a proper back-action evasion).
%are set to zero due to the case where the phonon population $n_{ph}>>1$ while the cavity is in its zero temperature limit. 
This allows for the mechanical spectrum to be computed straightforwardly  in the classical regime:
%\\
%\\
\begin{eqnarray}
%\begin{split}
        S_{\tilde{b}}(\omega) &=& |\chi_m(+\omega)|^2(1+|\bar{\Sigma}(+\omega)|^2)\Gamma_m N_{bath}   \label{Sb} ,\\
        S_{\tilde{b}^\dagger}(\omega)&=&|\chi_m(-\omega)|^2(1+|\bar{\Sigma}(-\omega)|^2)\Gamma_m N_{bath}   ,\label{SbDag}
%\end{split}
\end{eqnarray}
%\\
where $\chi_m(\omega), \bar{\Sigma}(\omega)$ are %the inverse self energy of the blue peak at $-\delta$ and $\bar{\Sigma(-\omega)}$ is the inverse self energy of the red peak at $\delta$.
given in Appendix \ref{theorystuff}. $N_{bath}= k_B T/(\hbar \Omega_m)$ is defined from the temperature $T$ the experiment is performed at. 
\\
\\Re-injecting the mechanical %spectrum 
solution into the optics %spectrum under 
equation, 
%the conditions that the cavity is in it's ground state and the mechanics are in the classical regime, the $n_{cav}$ noise, sideband asymmetry 
%and quantum fluctuations have negligible contributions. Injecting and simplifying
and neglecting cavity noise we get:
%%%%%% no, this is for below...
\begin{equation}
    S_{\delta\tilde{a} }(\omega)=C\frac{  \kappa_{tot}^2/4}{(\omega+\Delta)^2+\frac{\kappa_{tot}^2}{4}}\left[B_B \, S_{\tilde{b}}(\omega)+B_R \, S_{\tilde{b}^\dagger}(\omega)\right] ,\label{eqSa}
\end{equation}
\begin{equation}
    S_{\delta\tilde{a}^\dagger}(\omega)=C\frac{  \kappa_{tot}^2/4 }{(\omega-\Delta)^2+\frac{\kappa_{tot}^2}{4}}\left[B_B \, S_{\tilde{b}^\dagger}(\omega)+B_R \, S_{\tilde{b}}(\omega) \right]  , \label{eqSaDag}
\end{equation}
%\\
%%%% DEFINE Coop and ncav here!!!
with coefficients $B_B$ and $B_R$ given in Appendix  \ref{theorystuff}.
We have introduced %%%%%%%
$C$ the cooperativity:
\begin{equation}
 C=\frac{4 g_0^2 \,n_{cav}}{\kappa_{tot}\Gamma_m}
 \end{equation}
 defined from the (total) cavity population:
 \begin{equation}
     n_{cav}=\frac{\alpha_0^2 \,\kappa_{ext}}{\Omega_m^2+\kappa_{tot}^2/4} ,
 \end{equation}
%Experimentally here, the 
meaning that for us cooperativity is obtained from %considered via 
the combined power of both pumps, and  $n_{cav}$ is the total injected photon population. %of both drive fields.

 \begin{figure}
     \centering
     \includegraphics[width=1\linewidth]{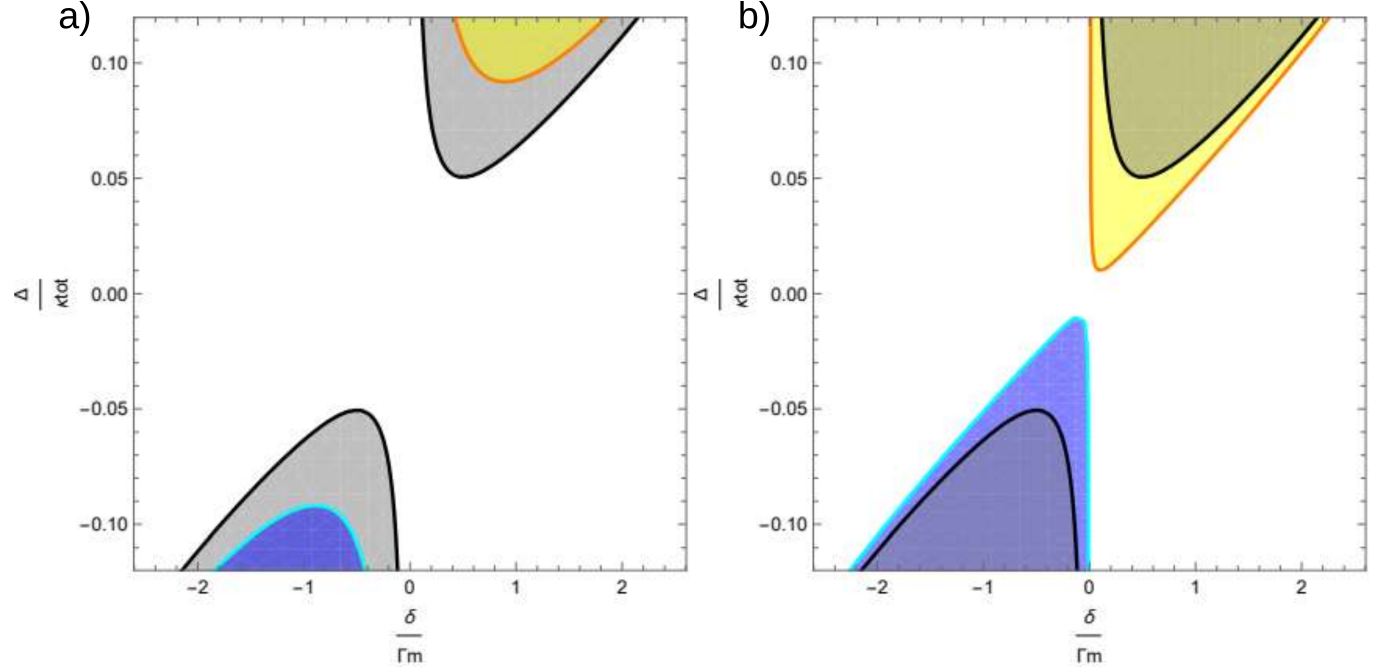}
     \caption{\label{fig:power} $\epsilon$-dependence. The black area displays the instability region under ideal conditions  Eq. (\ref{idealK}) with $C=10$. The orange and blue highlighted lobes display the self-driven dynamics with $S=0$ and: a) $\epsilon=-4\%$ showing that the instability region is reduced as %the effective dampening 
    the red pump contribution is increasing, b) $\epsilon=+4\%$, and the instability region grows due to increased blue power.}
 \end{figure}

The physical measurement is a field quadrature (for us a Voltage) related to the intracavity field through the input-output relation $\hat{V}(t)\propto \sqrt{\kappa_{ext}}\,[\hat{a}^\dagger(t)+\hat{a}(t)]$, see e.g. Ref. \cite{SchwabClerkSBAssymPRX}.
Neglecting the output photon background noise and sideband asymmetry, we can simply write $S_{\hat{V}}(\omega)\propto S_{\delta\hat{a}}(\omega-\omega_p)+S_{\delta\hat{a}^\dagger}(\omega+\omega_p)$; we also dropped the pumps-dependent term.
%%%
%%%this causes the spectrum to symmertise thus that the contribution of $S_{\delta\tilde{\alpha^\dagger}}(-\omega)\rightarrow0$. The full derivation and construction of the fitting plots can be found in the Appendix.
%%%%%%% This is incorrect! I rephrase it:
What is measured by our classical detector is the symmetrized part of the voltage spectrum. Looking at Eqs. (\ref{eqSa},\ref{eqSaDag}), we see that this simply leaves us with Eq. (\ref{eqSa}) shifted to $\omega_p$, with  $S_{\tilde{b}},S_{\tilde{b}^\dagger}$ replaced by Eqs. (\ref{Sb},\ref{SbDag}).
These look reasonably Lorentzian for any physical parameters we use in practice, and we numerically extract frequency position, width and area from the theoretical expressions (see Appendix \ref{theorystuff}).\\
%%% OK?

Diagonalizing the dynamical matrix and 
solving for parameters that collapse the real part to zero
% solving for where the real part collapses to zero
 gives the location of the instabilities, for a given set of tuning parameters \cite{KippenbergPRX}. %, these functions are solved numerically due higher order terms. 
This is performed numerically (see discussion in Appendix \ref{theorystuff}), constructing what we call an "instability map".
If we consider the case where $Q_0\rightarrow\infty$, $\epsilon\rightarrow0$, together with a sideband resolution $S\rightarrow0$, we are able to recover the "ideal" case analytical solution: % for the ideal case:
    \begin{equation}
        \frac{\delta \, \Delta }{\Gamma_m\, \kappa_{tot}} \left(\frac{C}{2}\right)=\left(\frac{1}{4}+\frac{\delta^2}{\Gamma_m^2}\right)\left(\frac{1}{4}+\frac{\Delta^2}{\kappa_{tot}^2}\right) . \label{idealK}
    \end{equation}
%%%Within these limits the results of the ideal case matching the theory detailed in the work of Kippenberg.
%%%%% badly said, you do NOT want to challenge Kippenberg.
This expression is exactly the one of Ref. \cite{KippenbergPRX}; note however the factor 2 difference in the definition of cooperativity that we use here.
%%% OK?
%\\
%as such this causes C in Eq.11 to inherit a factor of 2. $\Gamma_m$ is the linewidth of the mechanical resonator.\\
%\\

\begin{figure}
    \centering
    \includegraphics[width=0.7\linewidth]{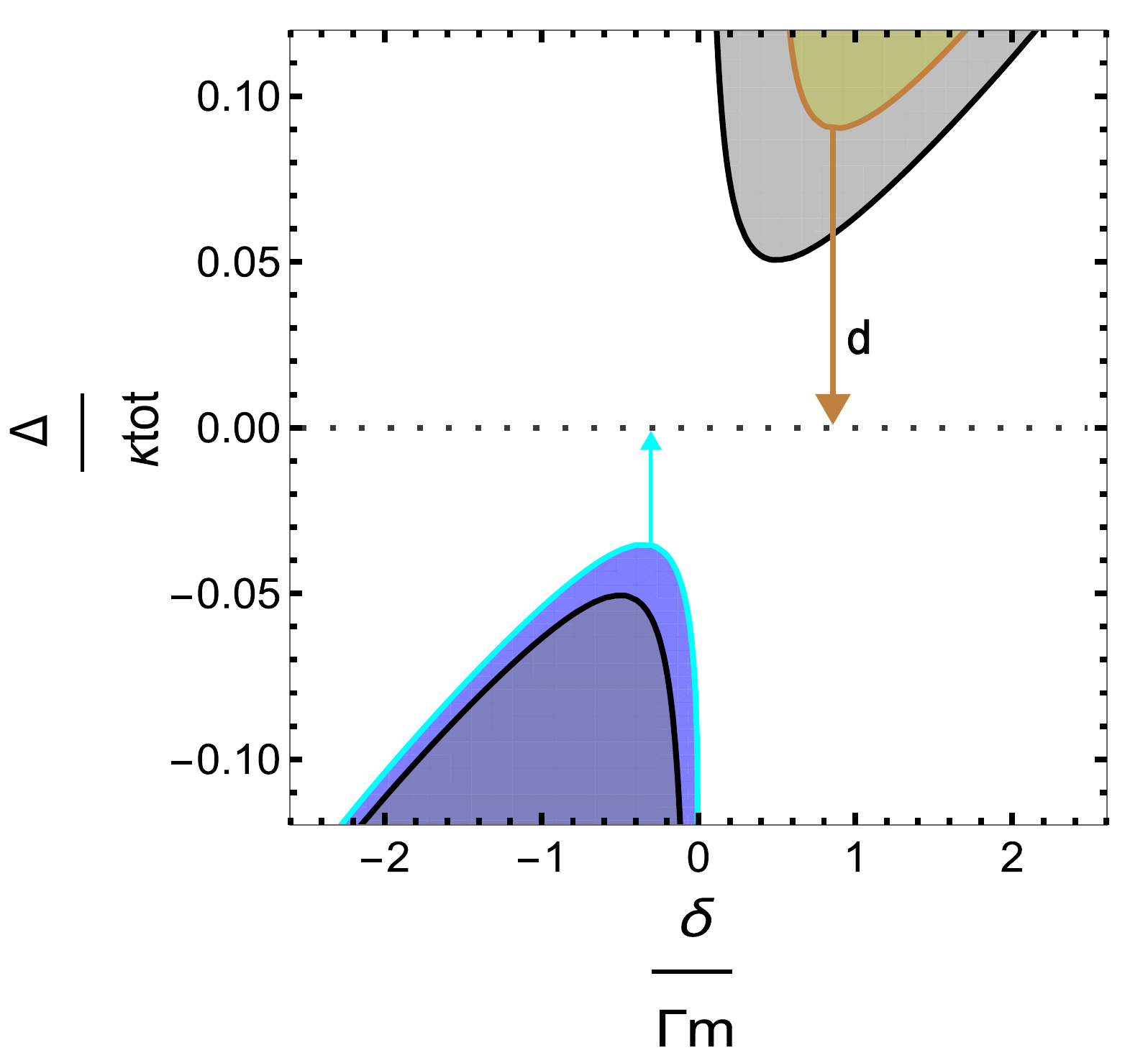}
    \caption{\label{fig:assym} $S$-dependence. The black area displays again the instability map under ideal conditions Eq. (\ref{idealK}) with $C=10$, the upper and lower zones are symmetric. The orange and blue highlighted lobes display the instability region should the system have an $S=0.45$, keeping $\epsilon=0$. As can be seen there is an %slight 
    asymmetry between the two zones, which can be measured by the distance $d$ to the $\Delta=0$ axis (arrows).}
\end{figure} 

Thus we can construct the theoretical plots comparing to this ideal case, in order to demonstrate the effect of the additional parameters $\epsilon, S$. 
We do so at a finite $Q_0=100\,000$, typical of experimental realizations. 
%%%%
It turns out that for reasonable values of $Q_0$ (typically larger than 1000), this parameter is not  especially relevant.
%%%%
Consider first the case $S=0$ but where $\epsilon\neq0$, therefore an imbalance in power (Fig. \ref{fig:power}).     
%%%%
%%%% this must be better explained!
We observe  an intuitive result: %is concluded, 
as the mechanical damping is 
increased with more red power
(Fig. \ref{fig:power} a), the borders of the instability map shift apart.
%increased 
%reduced %%%% !!!!! RQ: you DO NOT explain that red/blue increase decrease damping.... This is incomprehensible....
With an excess blue power (Fig. \ref{fig:power} b),  the borders of the instability map %reducing
close towards the center. % while when %amplification is increased
%the damping is increased with more red
% (Fig. \ref{fig:power} a) the borders of the instability map shift apart.
%, it is demonstrated experimentally that this reduction or growth is linear with $\epsilon$.
%%%% NO you do not demonstrate that here.
%\\
%\\
The effect grows with increasing cooperativity. Note in Fig. \ref{fig:power} the impact of an imbalance as small as $4\%$. 

\begin{figure}
    \centering
    \includegraphics[width=0.9\linewidth]{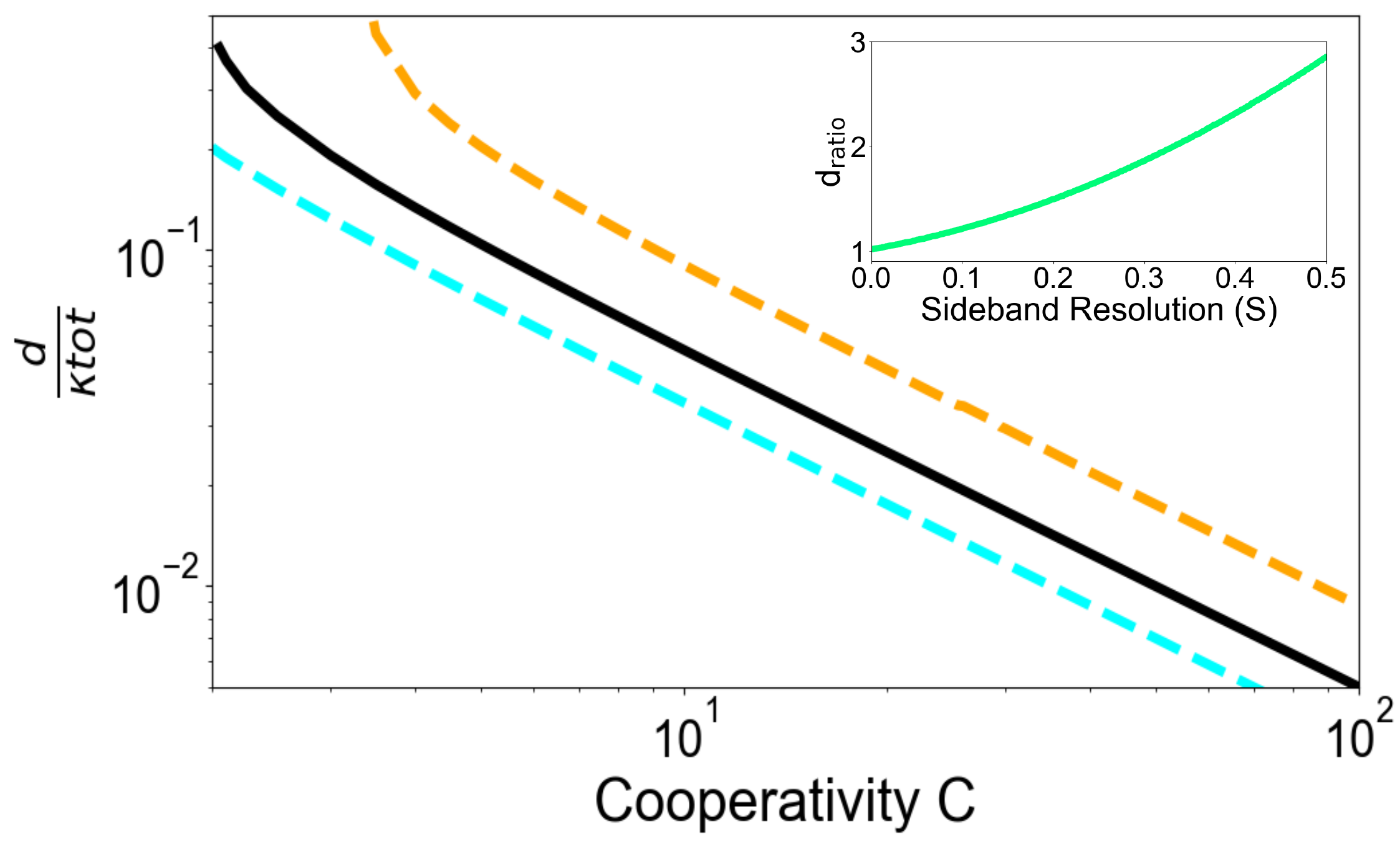}
    \caption{\label{fig:spacing} Main: the distance $d$ in relative units
    %%%% what units? Be more specific.
    between the tip of the instability region and the $\Delta=0$ axis. The log-log plot highlights a $1/C$ relationship. The black solid line shows that under ideal conditions $S=0$, the lobes maintain the same distance to the center. The orange and cyan dotted lines show the distance from the top and bottom lobes respectively with an $S=0.45$. These are different, and %, further it can be seen that 
    the orange dotted line has moved further from the ideal line than the cyan. Inset: $d$-ratio of top lobe distance over bottom one, as a function of $S$, illustrating the importance of the asymmetry effect as $S$ grows.}
\end{figure}

Considering the case where $S\neq0$ but keeping $\epsilon=0$, we observe 
%%%% YOU SHOULD INSIST!!!
a remarkable result:
that the map experiences an {\it asymmetry} %proportional
growing with %%%% You do NOT plot it as a function of S... same remark as the linear dependence with epsilon above...
 the $S$ parameter (Fig. \ref{fig:assym}, and inset Fig. \ref{fig:spacing}).
% It also turns out that $S$ and $Q_0$ have some interplay when $S \neq 0$, which {\it is visible} for typical device parameters.
 This has important implications experimentally, since it can bias an estimation of the actual center $\{ 0,0 \}$ of the map, the "sweet spot" for a back-action evading measurement.
 %%%%% OK???
%\\
To further illustrate this effect, we show how the distance $d$ of the instability lobes to the 
%center of the system
%0 of the
$\Delta=0$ axis changes with cooperativity (see arrows in Fig. \ref{fig:assym}). 
%showing that the asymmetry is a constant consideration and plays a significant role.
%%%% What?
%\\
In Fig. \ref{fig:spacing} as cooperativity grows, this spacing between the instability regions shrinks (noting that the trend is $1/C$), and we observe that as we leave the ideal case where $S=0$ (black line) %we see the effect and amplifying effects of 
there is a constant asymmetry. 
Typically, with $S=0.45$ the asymmetry between the lobe distances (cyan and orange) is as {\it large as a factor of 3}.
Besides, the gap between the two lobes is only $20~\%$ larger compared to the ideal case, which can be easily missed in an experiment.\\ %%%% Goooood!

%RQ: the y axis should read d/kappa_tot

\begin{figure}[h]
    \centering
    \includegraphics[width=1\linewidth]{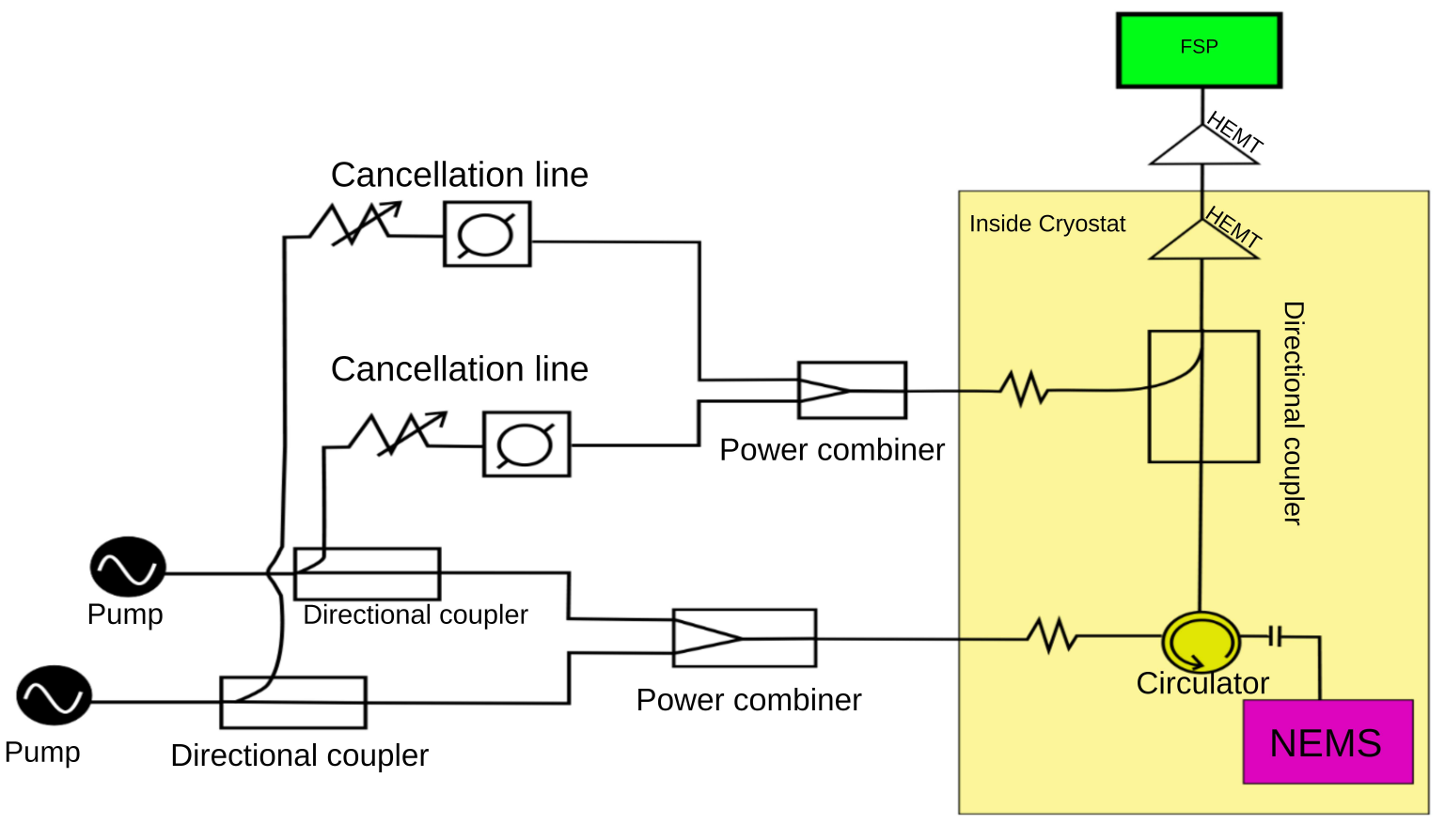}%, height=0.8\linewidth]{Setup_labeled.pdf}
    \caption{\label{fig:setup} Schematic of setup. 
    Two pump tones are injected in the cryostat (with 55 dB of attenuation). 
    %which output into a power divider, one line traveling to a power combiner before entering the cryostat while the other line is passed through an variable attenuator and phase shifter to establish a cancellation line, protecting the amplifiers.
They are split in an injection and a cancellation line.
%Upon entering the cryostat both the pumps input and opposition line are incident on a set of attenuators . The pump line passes through a circulator and are incident on the NEMS via a capacitive coupling. The reflected output signal is passed through a power combined 
The signal passes through two HEMTs (about 40 dB gain each) and is measured by a spectrum analyzer (FSP). }%directional coupler, mixed with the opposition line and amplified via a cryo/room temperature HEMTs before outputting to the FSP.}
\end{figure}

\section{Experiment and tuning procedure}

The Experimental setup consists of a movable aluminum membrane (NEMS mechanics), capacitively coupled to a superconducting aluminum $LC$ resonator (cavity).
The device is pumped via two microwave input lines with corresponding
cancellation 
lines, which inject the optical drive via a set of attenuators into a %BlueFors 
cyrostat regulated at 25 mK. 
The device is protected by a cryogenic circulator, 
and the output signal is sent through two sets of High Electron Mobility Transistor amplifiers (HEMTs) protected by the cancellation lines, and measured via a %Algient
Spectrum Analyser (FSP), see Fig. \ref{fig:setup}.
%%% YOU SHOULD EXPLAIN BLUE AND RED....

%%%% OK?
%%% Will need a rewrite once intro is done - Technically, a red detuned pump leads to optical damping, resulting in an effective linewidth $\Gamma_{ef\!f}$ larger than the original $\Gamma_m$. For a blue detuned pump, one obtains anti-damping and $\Gamma_{ef\!f}$ is smaller than $\Gamma_m$.

 \begin{figure}[h!]
     \centering
    \includegraphics[width=0.7\linewidth]{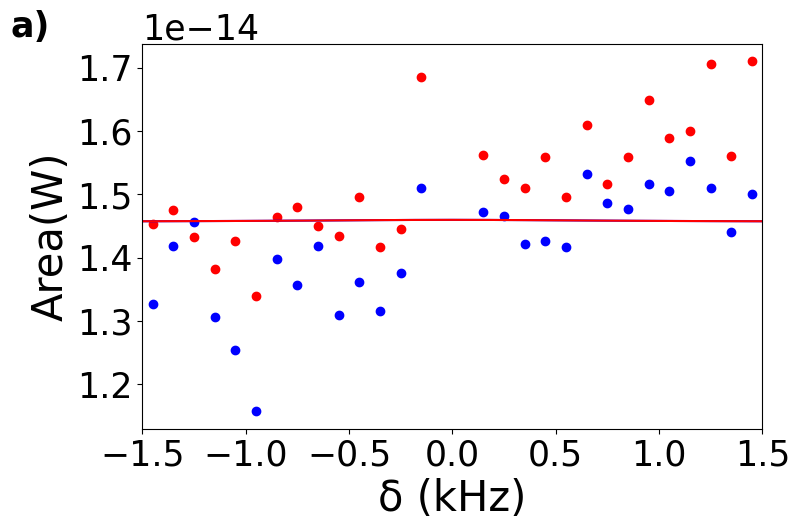}
    \includegraphics[width=0.7\linewidth]{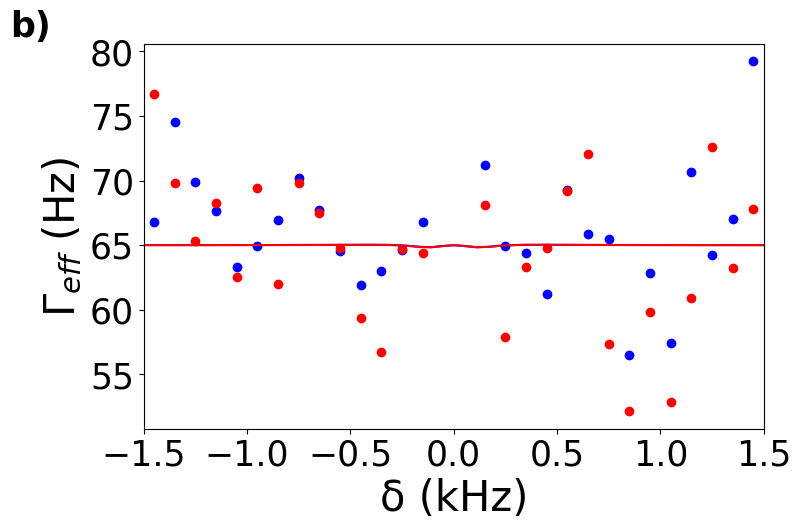}
    \\
    \includegraphics[width=0.7\linewidth]{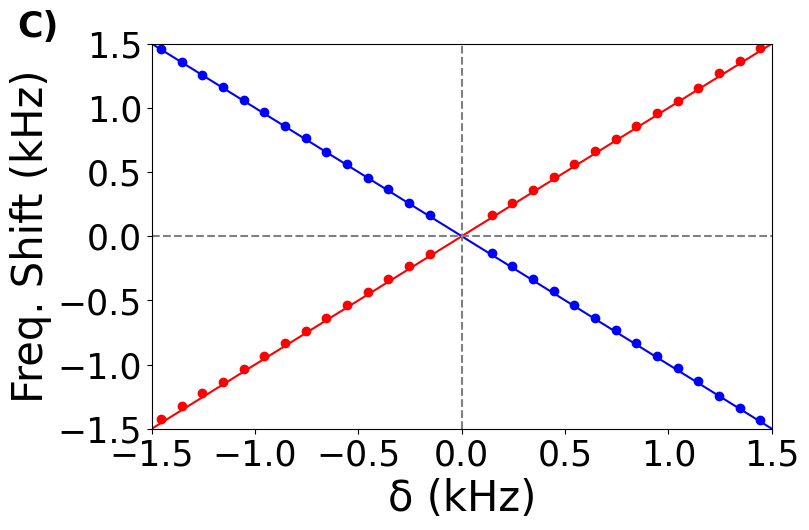}
     \caption{\label{fig:smalldelta}
     $\delta$ fine-tuning. $C=1$ measured sideband signals (blue dots for blue, and red dots for red). The data is obtained from Lorentzian fits of the peaks. The lines are from theory (numerics).
     %at  was used to construct, the blue dots correspond to the data points related to the data produced by the blue sideband and the red points are related to the data produced by the red sideband, this remains true for all future tuning plots. 
     a) The area of the peaks as a function of $\delta$ over $\pm$ 1.5 kHz, b) The measured linewidth $\Gamma_{ef\!f}$% across a $\delta$ of $\pm$ 1.5kHz
     , c) Relative peak positions as a function of $\delta$. The dashed lines indicate ideal tuning. }
\end{figure}
%\\
%\\

%Before performing the fine tuning, 
We first establish the baseline parameters of the device from independent cavity/mechanical characterization. This is done by a combination of conventional single-tone probe (for cavity), and red/blue pumping schemes (for mechanics) \cite{IlyaNJP}:
%which provide estimates of these characteristics which are necessary for estimating parameters such as Cooperativity:
\begin{itemize}
    \item $\omega_c=2\pi\cdot 6.4258$GHz is the cavity frequency,
    \item $\kappa_{tot}=2\pi\cdot 1.04$MHz is the cavity tot. damping rate,
    \item $\kappa_{ext}=2\pi\cdot 354$kHz is the cavity ext. coupling rate,
    \item $\Omega_m=2\pi\cdot8.473$MHz is the mechanical frequency,
    \item $\Gamma_m=2\pi\cdot65$Hz is the damping of the mechanics,
    \item $g_0=2\pi\cdot 60$Hz is the optomechanical coupling.
    %\item $S= 0.12$
\end{itemize}
%%%%
This makes our $S$ parameter about $0.1$ here.
At high cooperativity, the blue pump can drive the mechanics to self-oscillation, so the two-tone measurement sequence is initialized with the red pump switched on first. %this avoids the NEMS triggering a change of state into a self oscillating state by introducing a dampening effect due to the red pump. 
%Ideally this recovers a mechanical response with a linewidth $\Gamma_m$. 
Temporizations of a few seconds are used every time a pump setting is modified, in order to make sure the setup has reached its steady state. 

%To this end a fine tuning protocol was constructed which is discussed below. 
%%%% OK?

%%%
 \begin{figure}[t]
     \centering
    \includegraphics[width=0.7\linewidth]{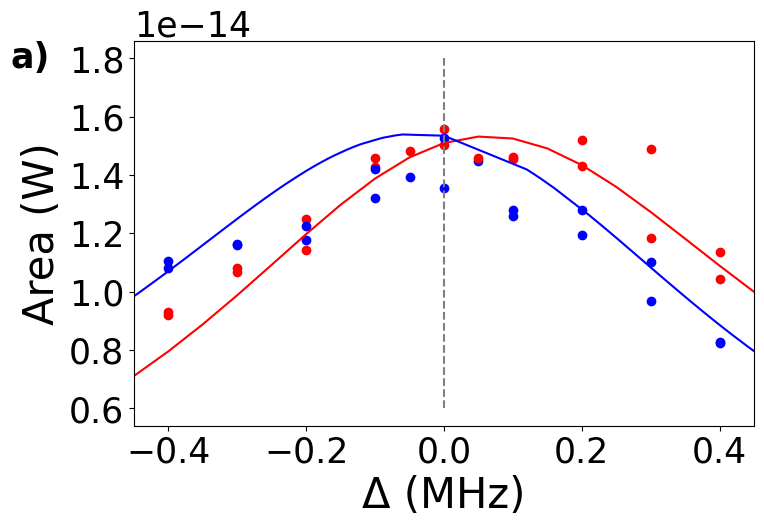}\\
    \includegraphics[width=0.7\linewidth]{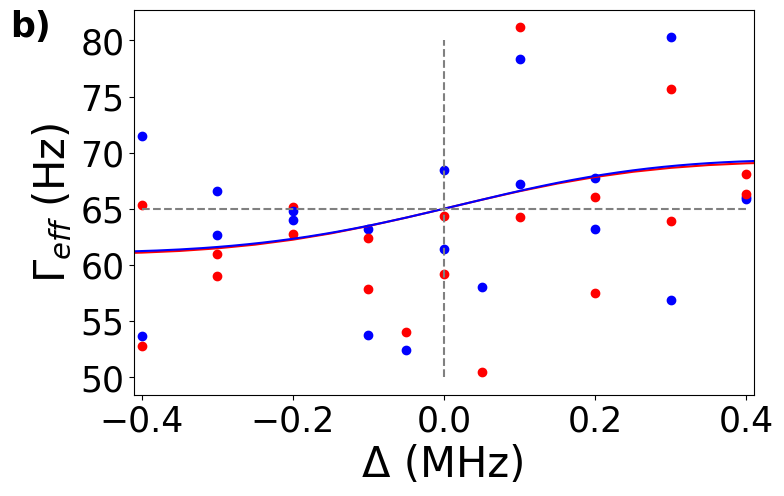}\\
    \includegraphics[width=0.7\linewidth]{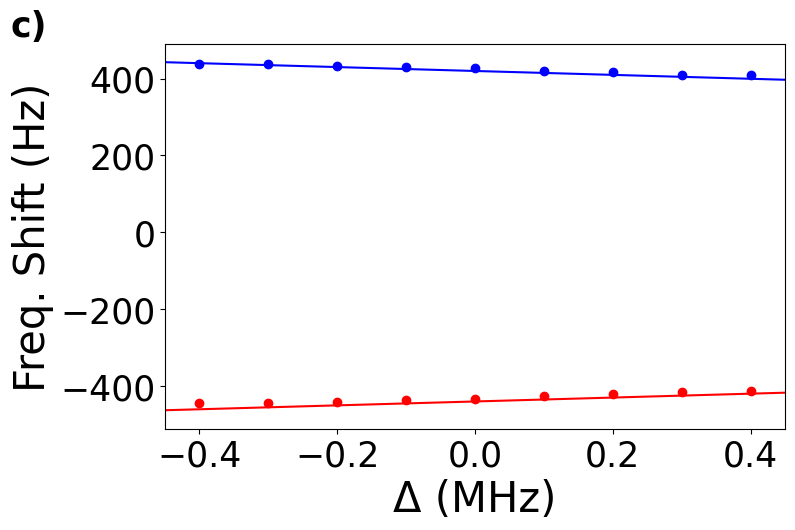}
     \caption{\label{fig:bigdelta}
     $\Delta$ fine-tuning. $C=1$ measured sideband signals (blue dots for blue, and red dots for red). The data is obtained from Lorentzian fits of the peaks. The lines are from theory (numerics).
     % The protocol to find the $\Delta=0$ was preformed using a similar method to FIG \ref{fig:smalldelta} with $C=1$ the red and blue dots are the experimental data for the red and blue pumps respectively, the red and blue solid lines are the predicted theory over 800 kHz.
     a) Area of the  peaks as a function of $\Delta$. The gray dashed line indicates the point where the areas match. b) Measured linewidth $\Gamma_{ef\!f}$; the dashed gray lines indicate where $\Gamma_{ef\!f}=\Gamma_m$.  c) Relative frequency position as a function of $\Delta$.}
 \end{figure}
 
The initial parameters for the two-tone drive are set using the original characterization given above. This provides a coarse tuning for the system, but due to the sensitivity of the experiment (which is increasing as cooperativity $C$ is increased), 
%becoming further sensitive at very large cooperativity $C$ as such mistunings are able to trigger self-oscillating states)
a subsequent fine-tuning procedure is required to reproducibly establish the measurement conditions.
To this end, the goal is to define the zeroes for the previously discussed detunings $\delta,\, \Delta$ and $\epsilon$.
%%%%  OK.
The tuning and measurement protocol was primarily performed at cooperativity $C= 1$, where it is easier. It was then repeated at higher cooperativities each time $C$ was increased. \\
%%%

\begin{figure}[h!]
    \centering
     \includegraphics[width=0.7\linewidth]{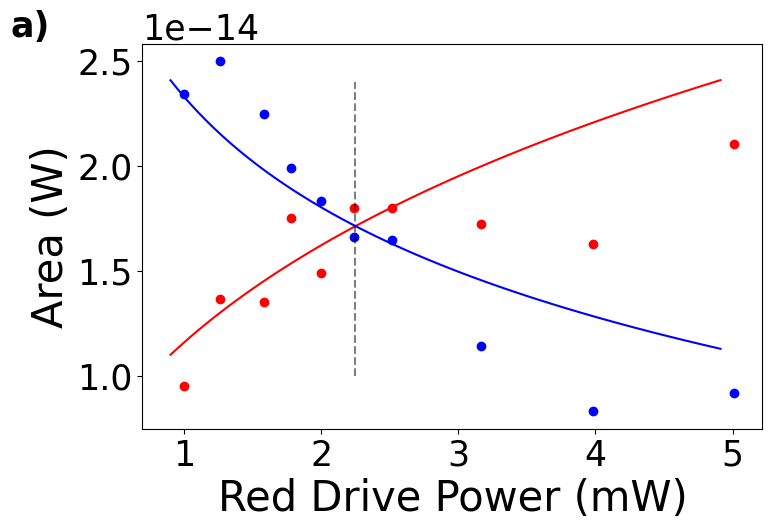}
    \includegraphics[width=0.7\linewidth]{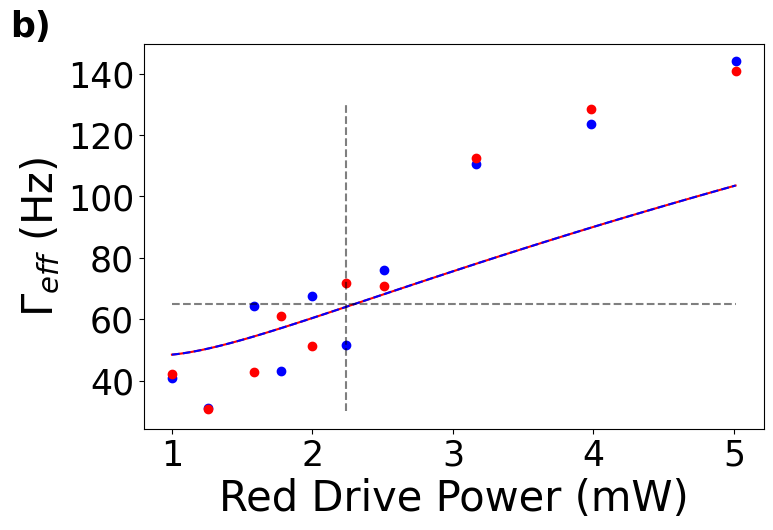}
    \includegraphics[width=0.7\linewidth]{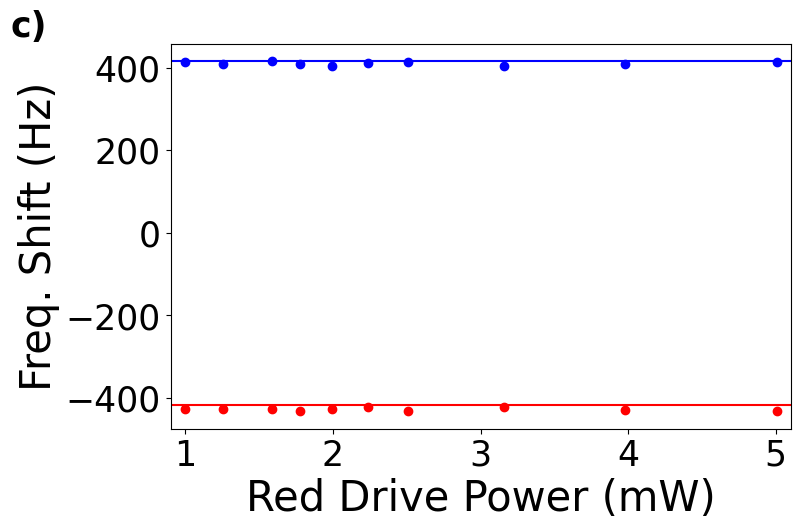}
    \caption{\label{fig:powertuning}
    $\epsilon$ fine-tuning. $C \sim 1$ measured sideband signals (blue dots for blue, and red dots for red). The data is obtained from Lorentzian fits of the peaks. The lines are from theory (numerics).
   % The protocol to find the $\epsilon=0$: 
    a) Area of the peaks as a function of injected red drive power. The dashed line indicates the point where $\epsilon=0$. b) Measured $\Gamma_{ef\!f}$.  The gray dashed lines indicate where $\Gamma_{ef\!f}=\Gamma_m$. c) Relative frequency position during the tuning process.}
\end{figure}
%\\

\begin{figure*}[t!]
    \centering
\includegraphics[width=0.75\linewidth]{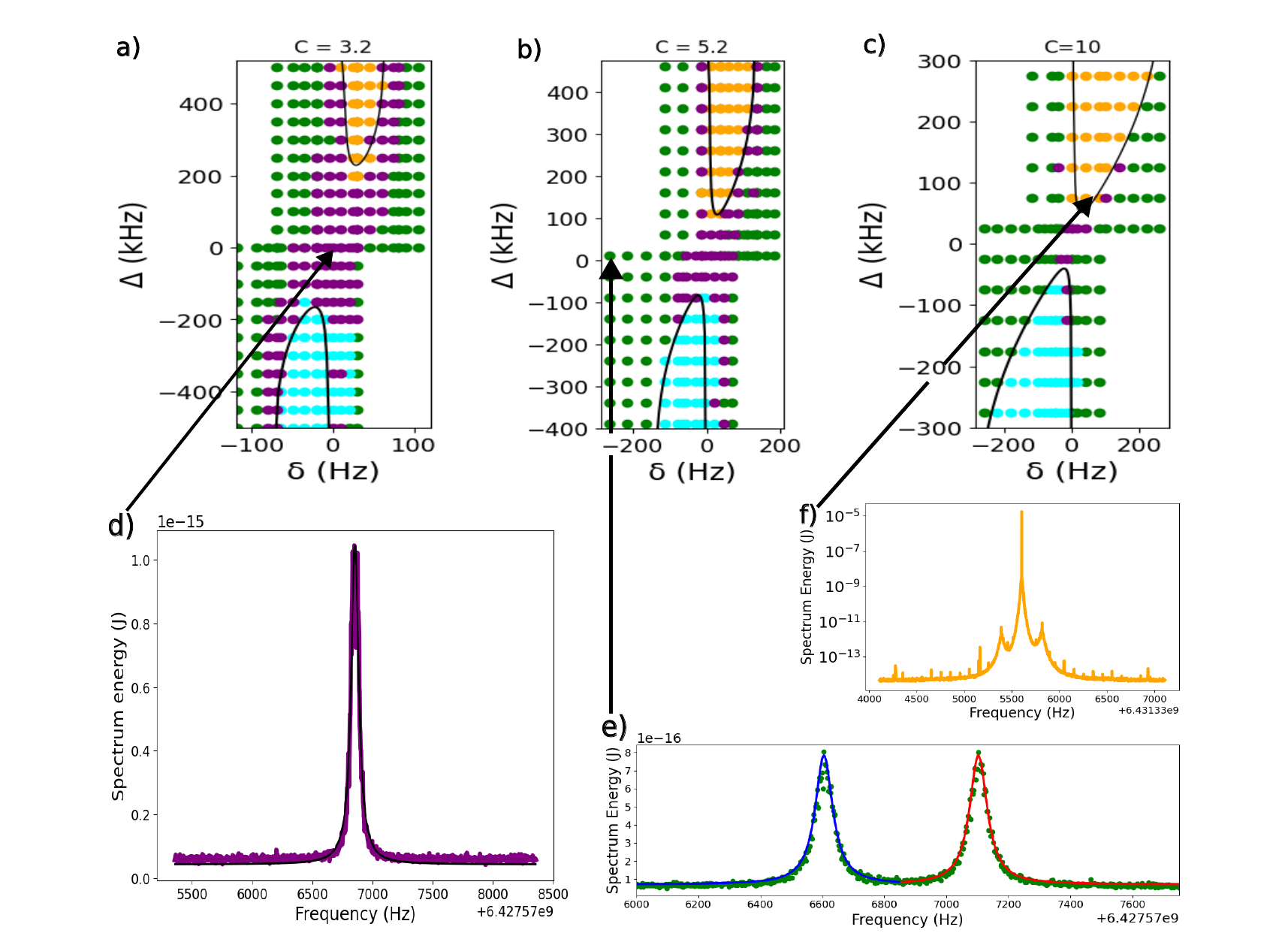}%\\
    \caption{\label{fig:map} a)-c) Instability maps obtained for cooperativity $C=3.2$, 5.2 and 10. (and $\epsilon = 0$). 
    In purple: zones with single (Brownian) Lorentzian peak (example in d, line is fit), in green two Lorentz peaks (inset e, "blue" and "red" sidebands with fits) and finally in orange/cyan the self-oscillating state (very large, and non-Lorentzian, see f). The black contours are theory, obtained for our system parameters.
    As cooperativity increases, the purple zone shrinks. Note the imperfections on the orange and cyan borders (see text for details).
    %At low cooperativity the increasing power causes the map to close proportionally with C, d) corresponds to the shape of the signal when the mechanics are projected properly on top of each other, e)shows the nature of the instabilities ,f) shows how the 2 signals can be separated. g)-i) At very high cooperativity the map continues to close but new purple points appear indicating the presence of new self-oscillating states, j) new self-oscillating states appear similar to a bessel function.
    }
\end{figure*}

The first procedure is to find the zero for the mechanical detuning ($\delta=0$). 
%%%
%Tuning the mechanical detuning is preformed by determining relative to $\Omega_m$ where the mechanical signals overlap. Using the initial settings from above we arrive to a good approximate, to finely tune $\delta$ we apply a shift imposing a shift in $\delta$ thus using the resulting frequency shift to find the crossing point, 
In the Brownian state with a finite $\delta$, two peaks are measured in the spectrum (see Fig. \ref{fig:map} e).
Extracting the relative frequency position of the two peaks as a function of $\delta$ produces Fig. \ref{fig:smalldelta} c).
The point at which they cross defines our $\delta=0$. %, the figure further shows that in the limits of C=1 the frequency shift is linear.
%%%%% 
%%%% why start with small delta?
As shown in Fig. \ref{fig:smalldelta} a) and b), $\delta$ alone does not noticeably affect damping (linewidth) and peak area. %; it simply fixes the positions of the peaks. 
This remains also true at higher cooperativities, with   however a small ($C$ dependent) optical spring %that shifts the peaks 
that can be resolved.
%%%% OK? YES.
Thanks to this negligible impact on linewidth and area, we can conveniently perform other tuning procedures while separating the blue and red sideband peaks with $\delta \sim$ a few $\Gamma_m$.
%Small delta is chosen as the first tuning parameter due to it having a negligible effect on the projected mechanics inducing a frequency shift, this makes it the simplest of the three tunings and is used as a convention so later tunings, the separated signals are relative to $\delta$. Separating the signals is done to allow better resolution to each pumps' dependancy on the tuning parameters.
On the other hand if $\delta$ is small compared to $\Gamma_m$, we eventually lose the resolution of the two peaks, see Fig. \ref{fig:map} d). 

The next tuning protocol aims at finding the %is to determine a relative 
$\Delta=0$ point, which can be defined as where the pumps $\omega_R,\omega_B$ are equidistant from the resonance frequency of the cavity $\omega_c$.
%, this is preformed by imposing a frequency shift on the pumps around $\omega_c$ and measuring the resultant area of the produced peaks. This tuning is preformed second as poor tuning in $\Delta$ can be apparently compensated via changing the input power, a however this will result in a system with both cavity and power tuning far from correct tuning.
% \\
% \\
In Fig. \ref{fig:bigdelta},  the pumps are positioned with %$\omega_m$ away from $\omega_c$ with an additional 
a detuning $\delta=400$ Hz in order to resolve blue and red peaks. 
%separating the peaks by 400Hz from $\delta=0$, the pumps are then sweept with a detuning relative to the cavity of 800kHz.
%%
Then, we sweep $\Delta$ and measure again area, width and relative position of the sidebands.
The exact $\Delta=0$ corresponds to the point where blue and red peak areas are equal, Fig. \ref{fig:bigdelta} a).
%%%% Mika's !!! XXX
Most of the feature seen is due to the cavity susceptibility appearing in Eq. (\ref{eqSa}).
%%%%
At this location, the linewidth of both peaks returns to $\Gamma_m$, see Fig. \ref{fig:bigdelta} b).
At $C=1$, the $\Gamma_{ef\!f}$ experiences a very minimal change with respect to $\Delta$; however while cooperativity increases, the magnitude of this effect becomes more and more prominent. 
%in both peaks and a linewidth equal to $\Gamma_m$, Fig.  shows that the area of the measured signal changes as a function of $\Delta$, this arises due to the sidebands seeing different transparencies into the cavity (this might need rewriting), the point at which the two theory plots in the area plot cross indicates the ideal tuning, it is further noted with a dashed line.
Note also the small $C$-dependent optical spring in Fig. \ref{fig:bigdelta} c). Both are roughly consistent with the slight $\delta$ imposed mistuning.  
 
 %at C=1 the theoretical shape is below the scatter of the data however, it increases in magnitude with cooperativity, though $\Gamma_{ef\!f}$ will always be equal to $\Gamma_m$ at $\Delta=0$. Though it is possible to see the effect of the optical spring in Fig. \ref{fig:bigdelta}, there is a small drift which is seen to grow with C, this cause the peaks to overlap far from $\delta=0$.
 %want to double check raw data that this is true
%\\

The final tuning protocol is to define when the injected power of the red and blue drives are equivalent.
This is performed by holding $|\alpha_B|$ at a fixed value, while sweeping the red power $|\alpha_R|$.
We again separate the blue and red sidebands with a $\delta=400$ Hz.
%such that $|\alpha_B|$ is equal to C=0.5 and varying the injected red power drive, this defines $\epsilon=0$ at C=1. The projected signals are detuned 400Hz from $\delta=0$,
$\epsilon=0$ is determined when the area of the two sidebands are equal, Fig. \ref{fig:powertuning} a), and equivalently when the linewidth matches $\Gamma_{ef\!f}=\Gamma_m$, \ref{fig:powertuning} b). The experiment was performed such that at that point, $C\approx 1$.
Fig. \ref{fig:powertuning} c) shows that the optical spring is essentially negligible.
%projected signals this mtaches resulting in $|\alpha_R|=|\alpha_B|$ as seen in Fig. \ref{fig:powertuning} which shows that 2.2 mW at the FSP the areas overlap, this can be further confirmed by seeing Fig. \ref{fig:powertuning} b) where one notices that where at $\epsilon=0$ tuned using the previous figure we see that $\Gamma_{ef\!f}=\Gamma_m$, Fig. \ref{fig:powertuning} c) shows that the power tuning simply alters the area and thus the linewidth of the signal and does not cause any frequency shifts.
%%%
Again features seen in area and linewidth become sharper at larger cooperativities. %\\

\section{Instability maps}

\begin{figure*}[t!]
    \centering
\includegraphics[width=0.75\linewidth]{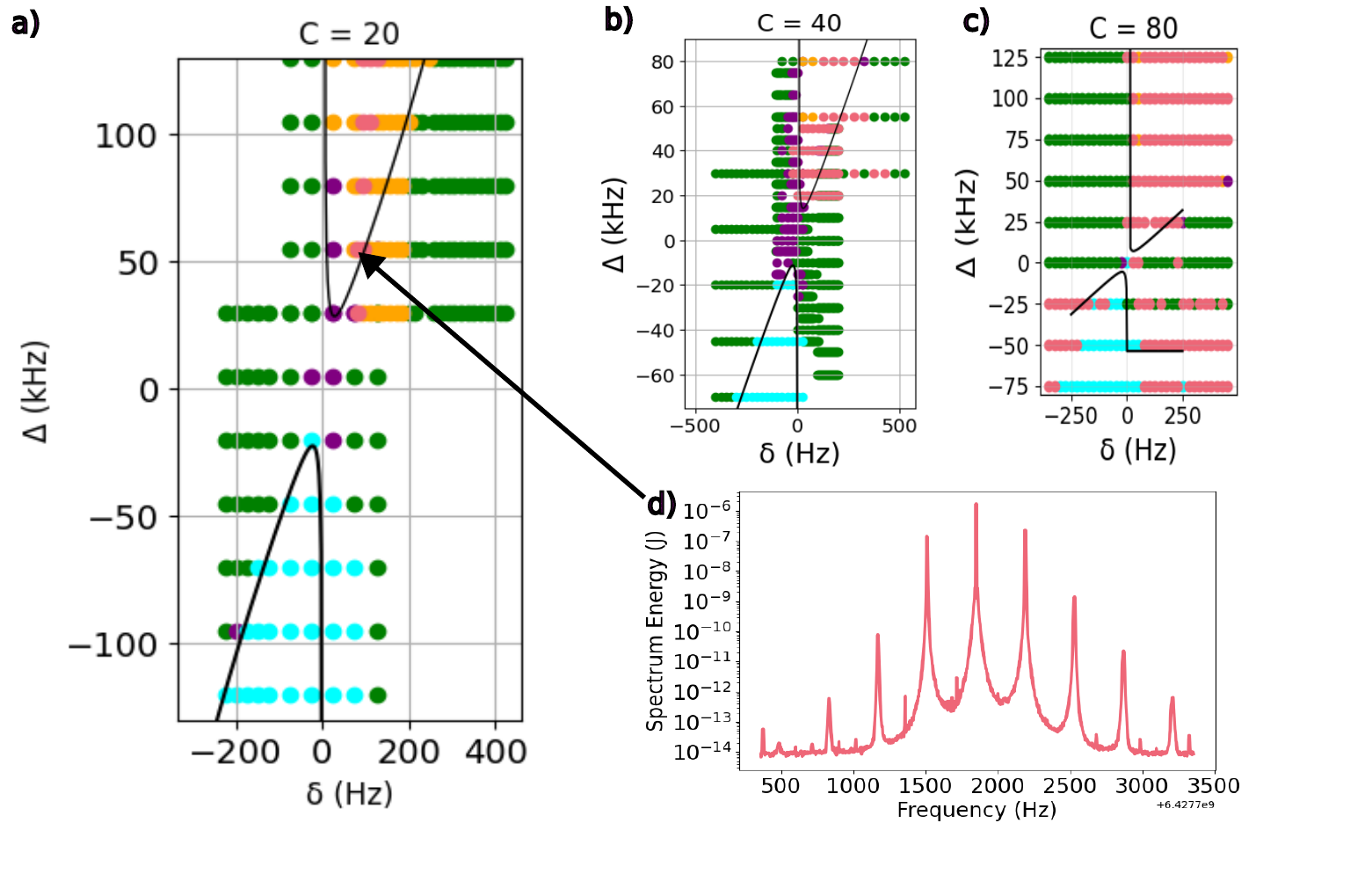}
%\vspace*{-1.5cm}
    \caption{\label{fig:mapBis} a-c) High cooperativity instability maps with $C=20$, 40 and 80 (and $\epsilon = 0$).
    The instability contour (theory in black) becomes more and more messy, with more and more irreproducibilities and hysteresis. In salmon, new self oscillating states appear displaying a mechanical comb (example in d). See text for details.
   % The set of produced instability maps, mapped from C=3.2 to C=80 a)-c) At low Cooperativity the increasing power cause the map to close proportionally with C, d) corresponds to the shape of the signal when the mechanics are projected properly on top of each other, e)shows the nature of the instabilities ,f) shows how the 2 signals can be separated. g)-i) At very high cooperativity the map continues to close but new purple points appear indicating the presence of new self-oscillating states, j) new self-oscillating states appear similar to a bessel function.
   }
\end{figure*}

In Fig. \ref{fig:map} we present instability maps at increasing cooperativity, ensuring $\epsilon=0$ for all configurations.
These are constructed while sweeping both $\delta$ and $\Delta$, and applying a threshold method:
If we detect a single "small" Lorentzian peak, as in Fig. \ref{fig:map} d), we display the region in purple.
This obviously depends on our resolution, and only right at the center is the measurement a single Lorentzian sideband. When two "small" peaks are detected (Fig. \ref{fig:map} e), the zone is green. Finally, if a "large" single peak is detected ("large" compared to "small" meaning about 100 times higher amplitude at least, see the $y$ axes), we display the map in orange (top) or cyan (bottom). Such a self-oscillating state is shown in Fig. \ref{fig:map} f).
    Note that, while we apply two drive tones, the triggered dynamics involves then a single peak (with some fine structure): the system somehow {\it synchronized} the response to the two pumps.
    This actually {\it does not happen} if one applies two distinct blue-detuned pumps, shifted by $\delta$ from each other.   
%, to construct the maps  are varied and the signals are averaged to obtain an accurate image of the mechanics, then the resultant peaks is sorted into one of 3 output types using a threshold method; single peak Fig. \ref{fig:map}.d) displays a measured signal peak, multi-peak Fig. \ref{fig:map}.f) shows that double peaks indicates where in the map do the two projected mechanics begin to separate and self-oscillating signal Fig. \ref{fig:map}.e) shows a measured self-oscillating peak which is decided to be self oscillating when the signal is 100 times the average height of a single peak.
In Figs. \ref{fig:map} a-c),  
the black lines show the theoretical instability contours, calculated with our system parameters. 
The agreement experiment versus theory is very good. However, we can already comment on the fact that the transition from Brownian to self-oscillation appears to be quite noisy; we will come back to that point at the end of the manuscript. Besides, a rather large hysteresis is seen, which becomes larger and larger as we increase cooperativity $C$. 
In order to understand this, one would need to provide a theoretical framework for the self-oscillating state that is triggered, as e.g. in Refs. \cite{AlexPRR,DylanPRR}.
Such a modeling could also interpret the heights of the peaks obtained in the self-oscillating state; note for instance that intriguingly, the main peak of Fig. \ref{fig:mapBis} d) is  {\it smaller} than the one of Fig. \ref{fig:map} f) while the applied power is twice larger. Saturation of the amplifiers cannot be  excluded with certainty.
%%%
These features are actually outside of our scope here.
We therefore make sure to always measure the contours switching from the Brownian state to the self-driven state; hysteresis will not be commented any further. \\
%
%It should be noted the noise around the theoretical plot, this noise shows a mix of the frequency noise (noise on $\omega_c,\Omega_m$), cooperativity noise (noise on $\Gamma_m,\epsilon$) and the hysteretic effects.

%What is the message of the maps
The same procedure has been followed up to $C=80$ which is remarkable, see Fig. \ref{fig:mapBis}. 
%%%
As cooperativity increases, the stable purple region, which contains the $\{\delta=0,\Delta=0\}$ "sweet spot" for a back-action-evading measurement shrinks.
At $C=80$, we are unable to resolve such a region, which illustrates our physical limit for the procedure.
%%%
Besides, at very high $C$ an intriguing feature appears: new self-oscillating states are triggered, displaying {\it  a comb structure} instead of a single peak (salmon color zone). It can present different shapes, and an example is shown in Fig. \ref{fig:mapBis} d). Such structured states seem to occur only in the top lobe of the unstable region, or in the hysteretic zones. The origin of this phenomenon remains mysterious. But the distance between the peaks seems to be related to $\delta$: it looks like the synchronization of the two pumps' response realized by the self-oscillation at low powers breaks down. This must obviously be linked to nonlinearities in the system, which have been kept out of our modeling \cite{DylanPRR,AlexPRR}. The small satellite peaks seen at lower powers in Fig. \ref{fig:map} f) might actually be precursors of the comb. \\
%Note that if we do pump with two blue tones above the parametric threshold, we do get two peaks close to $\omega_c$ (ore more precisely, two combs in the full spectrum). The fact that two pumps lead to a single mechanical oscillation state seems thus to be a specificity of the low power red$+$blue drive. \\
%Fig. \ref{fig:map}  demonstartes that the BAE can be tracked as C is ramped up, following as the instabilities close the map finding at C=80 the center is obscured by with instabilities. Further at very high C, it is revealed that new states are generated (the purple points on Fig. \ref{fig:map}.g-i), appearing similar to a bessel function (Fig \ref{fig:map}.j) suggesting that at high cooperativities the system is unable to {\it synchronize}, the spacing between the spikes in the comb are linked to both $\delta$ which heavily effects the spacing and further adjusted by $\Delta$ showing that it is subject to the optical spring effect. Uniquely, these new states appear in only one portion of the map up to the highest limit where a veritable "zoo" of new states is constructed. The power dependence on these new states suggests it is likely an interaction between the two tone scheme and kerr/duffing effect resulting in novel nonlinear states forming in the unstable region. 
%\\
%\\

\begin{figure}[h!]
    \centering
 \includegraphics[width=0.4\linewidth]{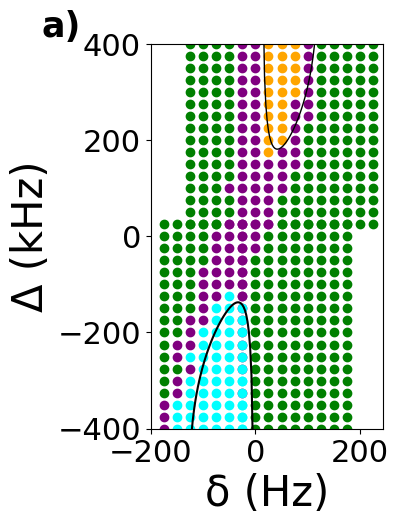}
\includegraphics[width=0.42\linewidth]{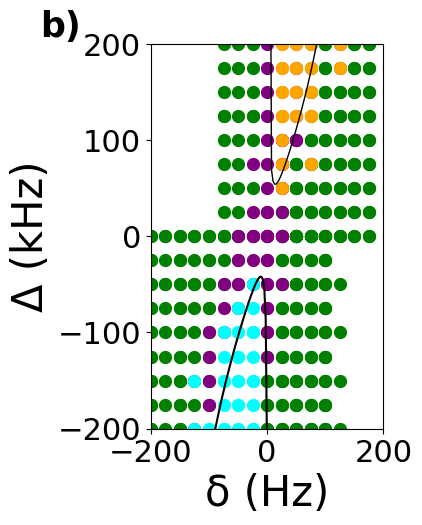}
\\
\includegraphics[width=0.9\linewidth]{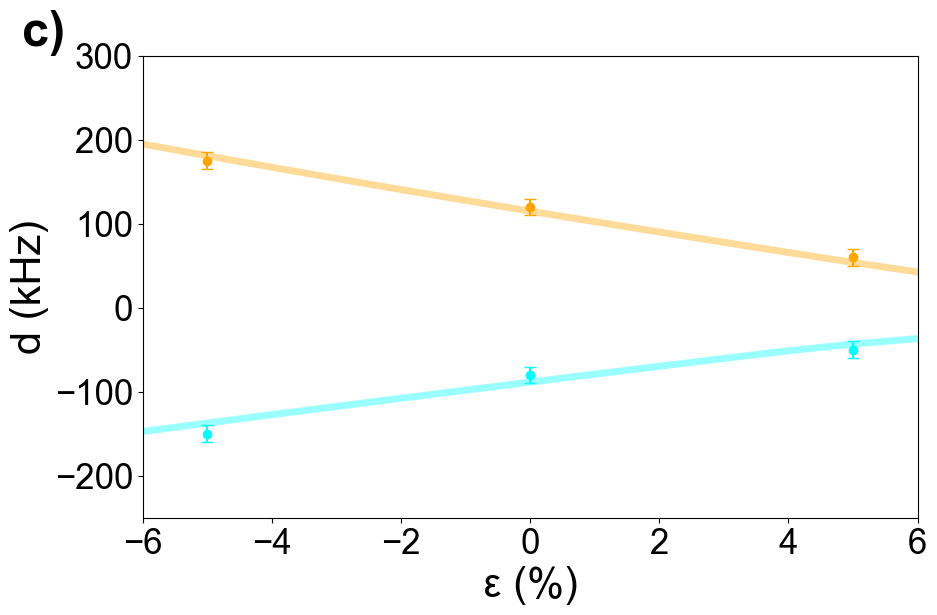}
    \caption{\label{fig:spacing_power} 
    %C=5.2 
    %The distance to the center point was measured for maps $\epsilon\pm5\%$ 
    a) Measured map at $\epsilon=-5\%$ and $C=5.2$. b) Same measurement with $\epsilon =+5\%$. As a comparison, Fig. \ref{fig:map} b) displays same cooperativity with $\epsilon=0$.
     c) Distance of the lobes to the axis. Orange dots show the top lobe, while cyan dots correspond to the lower one.
     %points for the top instability, the red solid line shows the theoretical plot, the blue dots show the points for the top instability, the blue solid line shows the theoretical plot, the theory and dots
     The lines are theory.}
\end{figure}

We further compare experiment to theory for nonzero $\epsilon$, see Fig. \ref{fig:spacing_power}.
We present instability maps for $\epsilon=\pm5\%$, and then plot the lobe distance $d$ to the $\Delta=0$ axis  as a function of $\epsilon$. The agreement is excellent, with the same degree of noise as on the previous maps. 
We point out again that even a pretty small power imbalance has a measurable effect. 
However, this is {\it not} our fundamental limiting factor here: we could somehow always tune the system properly, even at very large $C$. What seems to be the ultimate killer is the {\it intrinsic noise} of the system. \\
%Further, an $\epsilon$ of $\pm5\%$ was mapped at C=5.2 to observe what resulting deformation would occur,
%somewhat intuitively a positive $\epsilon$ causes the map to close as the amplification back-action can no longer be fully compensated, and visa versa the damping can no longer be fully compensated resulting in a wider map. The theory shows that the spacing remains linear the map, opening and closing by 50kHz.\\
%\\

\begin{figure}
    \centering
    \includegraphics[width=0.95\linewidth]{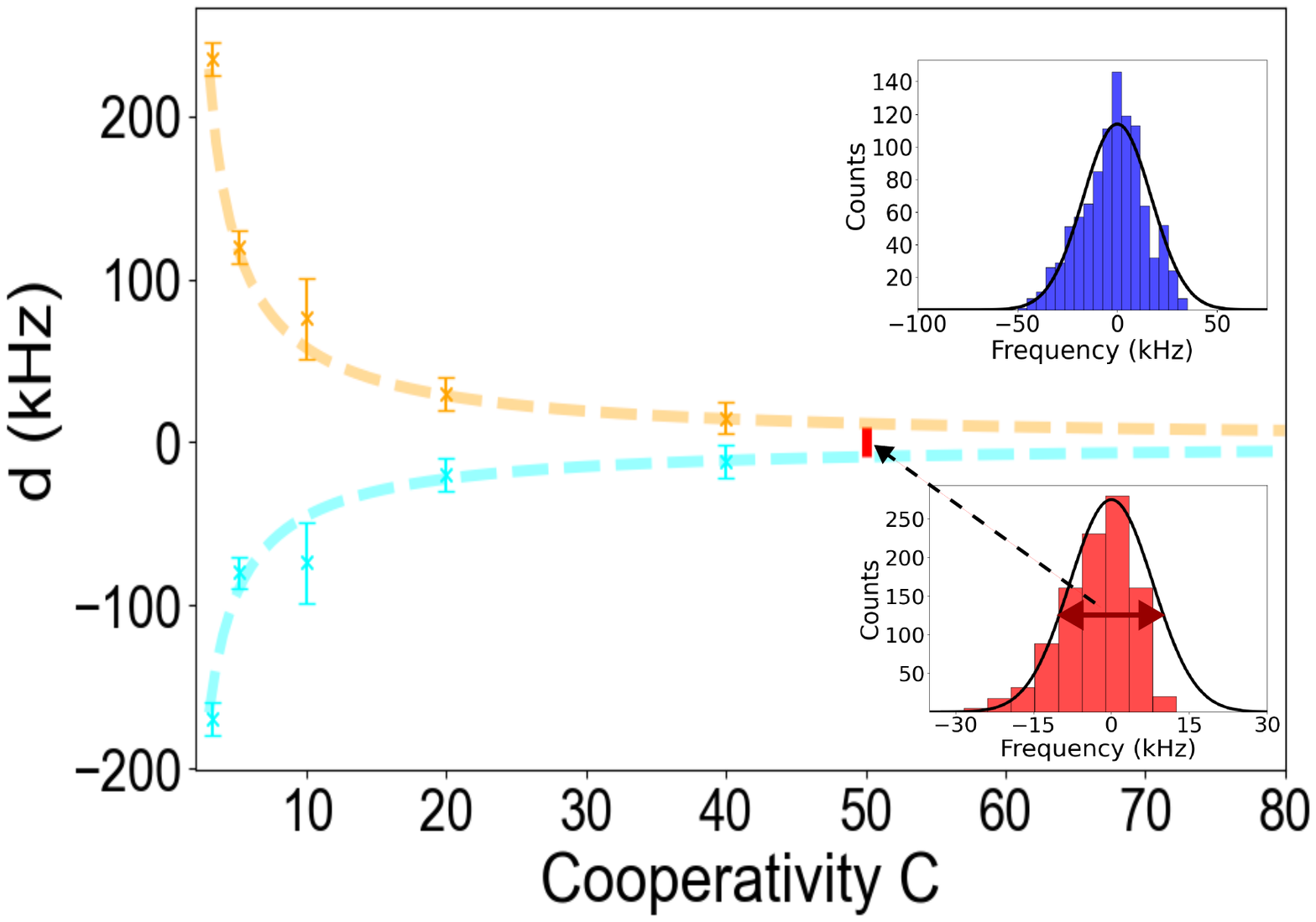}
    \caption{\label{fig:spacing_exp} 
    Main graph: Measured lobe distances $d$ extracted from the cooperativity maps of Figs. \ref{fig:map} and \ref{fig:mapBis}. The orange dots show the points for the top instability lobe, 
    %with the red solid line showing the theoretical plot, 
    and the cyan dots correspond to the bottom instability lobe.
    The dotted lines are theory.
    The error bars arise from the frequency step size of our acquisition in $\Delta$. The vertical red bar at $C=60$ corresponds to 20 kHz.
    %%%%
    %noise on $\Delta$, the error bars on C=10 are larger due to the map having a lower resolution at the tip of the lobes. 
    Bottom inset: Histogram of $\Delta$ noise while sweeping into the lower instability lobe, at $C=5.2$ (see text for details).
    Top inset: Similar measurement for noise in $\omega_c$  using a single-tone scheme (see text). }
\end{figure}

In order to study the physical limit of the two-tone drive, we thus plot in Fig. \ref{fig:spacing_exp} the top and bottom lobes distances  $d$ to the axis as a function of cooperativity $C$. 
It turns out that somewhere between $C=40$ and $80$, the fluctuations in tuning parameters are such that {\it the stability region in the center has essentially vanished}.
%%%
Noise in drive power seems to be unable to explain this phenomenon, as well as damping noise \cite{DylanPhD}: it can in principle never make the lower and upper lobes of the instability map overlap.
%%%
As well, the mechanical frequency noise in $\Omega_m$ is measured to be very small for our device (typically standard deviation of order 1 Hz); which also confirms that our measurement chain is more stable than tat.
%%% OK? 
Therefore, it is likely that the main source of instability is {\it cavity resonance frequency $\omega_c$ noise}.
% We evaluate distance between lobes $d$ (Fig. \ref{fig:spacing_exp}), comparing the total distance between the blue and red theory plots to the BAE noise along $\Delta$ find that the limit at which the maps close is when $C = 60$, where the frequency noise on $\Delta$ causes the instabilities to obscure the center.
%%%
%%% XXX!

This noise is estimated by making use of a simple technique inspired from Ref. \cite{nano}, which deals with frequency-noise in Duffing mechanical resonators.
For a given cooperativity $C$, we position the pumps at a fixed $\delta$ and $\Delta$ in the stable region, and then sweep slowly  
%$\delta=\pm10$ (this is to account for the the tip of the lobe being slightly offset from $\delta=0$) and 
$\Delta$  until the instability is reached. 
A histogram is built by repeating  the procedure about 2000 times.
Different settings have been tested, leading to essentially equivalent results.
%%%%%
Note that if the sweep is too fast, the bifurcation process can be activated too soon; one needs to have a slow enough process to be in an adiabatic limit. To some extent, this could explain part of the irregularities seen in Figs. \ref{fig:map} and \ref{fig:mapBis}. We thus verified that changing the ramp speed typically by a factor 2 does not alter much the histogram measurement. Indeed for a $1/f$-type frequency noise, we would expect a variance varying only logarithmically $\ln(t_{max}/t_{min})$, with $t_{min}$ the time needed for a single acquisition (about a couple minutes), and $t_{max}$ the total time span (here slightly over 4 hours) \cite{nano}.
A typical measurement is shown in bottom inset Fig. \ref{fig:spacing_exp}.
%%%%%%%%%%%%%%%%%%%%%%%%%%%
We obtain an Allan standard deviation of order 10 kHz by means of a Gaussian fit. And 20 kHz is precisely the spacing between the lobes obtained at $C=60$, see Fig. \ref{fig:spacing_exp} main plot (the vertical red bar).
%, a standard deviation is built finding a $\sigma=8$kHz which equates to a linewidth of 18kHz shown in the bottom insert of Fig. \ref{fig:spacing_exp}, 
%%%%

However, 
an oddity with these statistics is that the built histogram shows an asymmetric "crunched" shape outside the instability zone (on the right, in the lower inset Fig. \ref{fig:spacing_exp}), with a tail that extends far inside (on the left, not visible at this scale). The origin of this is unknown, but might be a remaining finite-speed sweep signature.
%%%
%%%
%, this shape and linewidth was found to be consistent between the lobes and cooperativities.
%%%%%%%
We compared this statistics with the equivalent one built using a standard (blue-detuned) single-tone drive, using a similar methodology \cite{DylanPhD}.
The cavity noise histogram 
is displayed 
 in the top inset of Fig. \ref{fig:spacing_exp}, and looks much more symmetric.
 The Allan standard deviation measured over 10 hours is about 20 kHz (see Gaussian fit).
 %which was found to have a $\sigma$ of 40kHz, this was determined using the technique developed in the work of Dylan et al{ref Dylan blue map}. 
 This number is in reasonable agreement with values reported for microfabricated cavities at milliKelvin temperatures \cite{niepce} (Fig. 1, panels E and F: in the 1 - 20 kHz range depending on material). 
 We see that the cavity frequency noise inferred from these two procedures are of the same order. 
 To go beyond this qualitative assessment, a complete theory of the noise impact on the bifurcation process is required.
 %
 %Showing that the intrinstic cavity noise and $\Delta$ noise are of the same order of magnitude. Though there is a discrepancy between the noise measurements, as such further work developing a quantify theory for the noise in BAE is required to fully explain the features.
% might need rewriting

%%%%%%% HEEEEEEEEEEEEEEEEEEERE
%\textbf{Check list -> VS code}
%\begin{itemize}
    % \item     make map clearer ; split in 2!
   %\item     make orange/cyan arrow with a black "d" label in the first map.
    %\item: on the diagram: attenuators HEMT are missing...
    % \item Fix axis size!
    % \item HORIZONTAL dash for tuning Delta Gamma 
    % \item Orange and cyan points + Thicker lines
    % \item add the kappa ext value!
    % \item put names on components of the setup, like attenuators, dir coupler, cancellation line, circulator, HEMT
%\end{itemize}
%%%%%%% HEEEEEEEEEEEEEEEEEEERE

\section{Conclusion}

%We expand previous works on the two-tone optomechanical instability, 
We present the full modeling of the "blue+red detuned" two-tone instability that takes into account {\it all} mistunings ($\delta, \Delta$ and $\epsilon$) {\it and} imperfections ($Q_0, S$), and compare it with experimental data. 
The theory is shown to match the experiment very well, with measurements performed up to an extremely high cooperativity $C$ of almost 100.
The capability of making the protocol work that well builds on a fine-tuning procedure which is presented, again with theoretical calculations.
% presenting and utilizing a fine tuning protocol allowing for tuning up to the highest Cooperativites. 

Power imbalance $\epsilon$ and sideband resolved parameter $S$, which had not been addressed so far, are shown to be particularly relevant.
Minute errors in power balance ruin the tuning, while a finite sideband resolved parameter can be a problem to define the "sweet spot" sought for back-action-evading experiments (the actual frequency tuning $\{0,0\}$ center).
%%%%%
%It is noted that the parameter with greatest effect on the tuning arises from the sideband resolution as in the case where $S\neq0$ an asymmertry is found the structure of the instabilities along the map.
It is also shown that the instability mapping can be tracked up to the physical limit, where the bottleneck seems to be the frequency noise on $\omega_c$, the cavity resonance position. 
%%%
Conversely, the mechanical element's inherent resonance frequency noise, and our generator's phase noise seem to play no role.
%%%
Having identified this constraining phenomenon is an important step forward, if one is to overcome it: one can aim at producing more stable microwave devices (with less internal two-level systems responsible for this noise \cite{burnett}), or if this happens to be almost impossible one can develop more involved schemes (like e.g. tracking the cavity frequency with a feed-back technique).
%{Therefore, the ultimate limit for stabilizing BAE at high powers is the cavity noise, which places a fixed upper limit until the center is obscured. (I want to move this here but it is late so I will work this in better tomorrow)} 
%%%%

An actual experiment being never ideal, proper tuning and characterization requires to take into account all of these problems. As well, since the measured signal is proportional to cooperativity $C$, improved signal-to-noise requires specifically high power stability.
For all of these reasons, we believe that our results are particularly relevant to the community, with implications also for other two-tone similar schemes \cite{lehnert,mikastuff}. 

There are still some open questions for future investigations, as at high $C$ the self-oscillating state triggered by the instability demonstrates new unexpected features: one can observe mechanical comb spectra with a spacing of order $\delta$, which suggest that the synchronization realized between the responses to the two pump drives at lower powers is destroyed. 
The heights of the self-oscillating peaks also deserve to be modeled. These theoretical developments will certainly require non-linear effects to be taken into account \cite{DylanPRR,AlexPRR}.
As well, a proper model taking into account the $1/f$ frequency noise quantitatively is still lacking.

%instability transitions into new states. 

%\begin{itemize}
%    \item We present the Full theory and demonstrate it experimentally
%    \item The Theory matches the experiment very well up to C=80
%    \item Present Fine tuning procedure
%    \item Importance of the S -> asymmerty
%    \item Show the physical limit limiting to omega c noise
%    \item Open questions High C causes instability to transition to new states
%\end{itemize}

%\newpage
%\vspace*{1cm}
(\dag) Corresponding Author: eddy.collin@neel.cnrs.fr

\begin{acknowledgements}

The Authors ave very grateful to Mika Sillanp\"a\"a who provided the sample used in the experiment, and helped us with many fruitful discussions.
We acknowledge the facilities and technical support of Otaniemi research infrastructure for Micro and Nanotechnologies (OtaNano).
The work was supported by the Research Council of Finland through the Finnish Quantum Flagship project (project number 358877, Aalto University).
L.M.d.L acknowledges funding from the Strategic Research Council at the Academy of Finland (Grant No. 338565).
 We also acknowledge valuable support from the N\'eel laboratory and the 
 European Microkelvin Platform (EMP) consortium, 
visit: \underline{https://emplatform.eu/}.

\end{acknowledgements}

\section{Data availability}

The data used for this manuscript (as well as a Mathematica\textsuperscript{\textregistered}$\,$ code) is available at the following URL: 
\small{\underline{https://cloud.neel.cnrs.fr/index.php/s/CnnYPKn8XHYZgXa}}.
% It corresponds to all calculations presented here.

\appendix

\section{Dynamical matrix and solving}
\label{theorystuff}

We present in this Appendix a summary of the theory built around the full Hamiltonian Eq. (\ref{hamilton}). For details, see the Mathematica\textsuperscript{\textregistered}$\,$ code.
The found mechanics' dynamical matrix involves the quantities:
\begin{eqnarray}
  &&  \gamma(\omega) = - C\, K \, \kappa_{tot} \times \\
  &&  \left( |B_{-}|^2 \frac{\kappa_{tot}^2/4+(\omega+\Delta)^2}{\kappa_{tot}^2/4} -|B_{+}|^2 \frac{\kappa_{tot}^2/4+(\omega-\Delta)^2}{\kappa_{tot}^2/4} \right) , \nonumber \\
 &&   \lambda(\omega)  =  - C\, K \, \kappa_{tot} \times   \\
 && \left( |B_{-}|^2 \frac{(\Delta-\omega)}{\kappa_{tot}}\frac{\kappa_{tot}^2/4+(\omega+\Delta)^2}{\kappa_{tot}^2/4} \nonumber \right. \\
 &&\left. \,\,\,\,\,\,\,\,\,\,\,\,\,\,\,\,\,\,\,\,\,\,\,\,\,\,\,\,  +|B_{+}|^2 \frac{(\Delta+\omega)}{\kappa_{tot}}\frac{\kappa_{tot}^2/4+(\omega-\Delta)^2}{\kappa_{tot}^2/4} \right) , \nonumber  \\
 &&   \Sigma(\omega)  =  - C\, K \, \Delta \times \\
 &&\left(2 e^{-2i \delta \phi} B_{-}B_{+}^*  \left[ \frac{\omega \,\kappa_{tot}+i\, (\omega^2-\Delta^2-\kappa_{tot}^2/4)}{\kappa_{tot}^2/4} \right]\right) , \nonumber
\end{eqnarray}
having introduced the detuning-dependent constants ($\omega_m = \Omega_m+\delta$):
\begin{eqnarray}
    |B_{-}|^2 & =& (1+ \epsilon)^2 \, \frac{(\omega_m-\Delta)^2+\kappa_{tot}^2/4}{\Omega_m^2 + \kappa_{tot}^2/4} , \\
    |B_{+}|^2 & = & (1 - \epsilon)^2 \, \frac{(\omega_m+\Delta)^2+\kappa_{tot}^2/4}{\Omega_m^2 + \kappa_{tot}^2/4} , \\
    B_{-}B_{+}^* & = & (1+ \epsilon)(1- \epsilon)  \times \\
    && \frac{(\Delta^2-\omega_m^2+\kappa_{tot}^2/4)+ i \,\omega_m \,\kappa_{tot}}{\Omega_m^2 + \kappa_{tot}^2/4} , \nonumber
\end{eqnarray}
and:
\begin{eqnarray}
    K & = &\frac{\Gamma_m \kappa_{tot} \, (\Omega_m^2 + \kappa_{tot}^2/4)^2 \, \kappa_{tot}^2/4}{8 \,[(\Delta^2-\omega_m^2-\kappa_{tot}^2/4)^2 + (\Delta \kappa_{tot})^2]^2} \nonumber \\
    & & \!\!\!\! \times \frac{[\kappa_{tot}^2/4+(\Delta+\omega_m)^2][\kappa_{tot}^2/4+(\Delta-\omega_m)^2]}{[\kappa_{tot}^2/4+(\Delta+\omega )^2][\kappa_{tot}^2/4+(\Delta-\omega )^2]}  .
\end{eqnarray}
Turning to the detected signal, we have used in the core of the manuscript:
\begin{eqnarray}
&&\!\!\!\!\!\!\!\!\!\!\!\!\!\!\!\!B_B  =   K' \, |B_{+}|^2 , \\
&&\!\!\!\!\!\!\!\!\!\!\!\!\!\!\!\!B_R  =   K' \, |B_{-}|^2 , \\
&&\!\!\!\!\!\!\!\!\!\!\!\!\!\!\!\!K'   =    K \, \frac{[\kappa_{tot}^2/4+(\Delta+\omega )^2][\kappa_{tot}^2/4+(\Delta-\omega )^2]}{(\kappa_{tot}^2/4)^2}   .
\end{eqnarray}
All of these expressions are exact, within a double rotating wave transform. \\

We can first look for the instability. We diagonalize the dynamical matrix and look for the zero of the real part of one of the eigenvalues.
This can be cast into the equation:
\begin{eqnarray}
    \frac{\Gamma_m}{2} + \frac{[\gamma(\omega)+\gamma(-\omega)]/2}{2}-\mbox{Re}\left[\sqrt{R(\omega)}\, \right] & = & 0,
\end{eqnarray}
with: 
\begin{eqnarray}
&&   R(\omega)  =  \left(\frac{\gamma(\omega)-\gamma(-\omega)}{4}\right)^2 -\left(\delta+\frac{\lambda(\omega)+\lambda(-\omega)}{2}\right)^2 \\
   && + \Sigma(\omega)\Sigma^*(-\omega) -2 i \,\left(\frac{\gamma(\omega)-\gamma(-\omega)}{4}\right) \left(\delta+\frac{\lambda(\omega)+\lambda(-\omega)}{2}\right)  . \nonumber 
\end{eqnarray}
This expression is rather involved.
%%%%
Note that it depends on $\omega$, which means that in practice the bifurcation criterion towards instability could depend on the position in frequency within the sideband peak. In fact, this dependence is essentially nonexistent for $\Gamma_m \ll \kappa_{tot}$; we can thus technically take $\omega \sim 0$ in the above expressions.
%%%%
Moreover, it turns out that $R(\omega)$ depends quite strongly on the sideband resolution parameter $S=\kappa_{tot}/\Omega_m$; while the mechanical quality factor $Q=\Omega_m/\Gamma_m$ does not play much role here. 
%%%
Only in the limit $S \rightarrow 0^+$ do we recover the simple expression of Ref. \cite{KippenbergPRX}. \\

Next, we can invert the dynamical matrix to find out the mechanical susceptibility matrix. This one writes:
\begin{equation}
\begin{bmatrix}
   \chi_m(\omega) & \bar{\Sigma}(\omega) \, \chi_m(\omega)\\
   \bar{\Sigma}^*(-\omega) \, \chi_m^*(-\omega) &  \chi_m^*(-\omega) 
\end{bmatrix} ,
\end{equation}
having defined:
\begin{eqnarray}
&&\!\!\!\!\!\!\!\! \bar{\Sigma}(\omega)  = \frac{-\Sigma(\omega)}{\Gamma_m/2 + \gamma(-\omega)/2 + i\, (\delta+\lambda(-\omega) - \omega) } , \\
&&\!\!\!\!\!\!\!\! \chi_m(\omega) =  \left[  \Gamma_m/2 + \gamma(-\omega)/2 + i\, (\delta+\lambda(-\omega) - \omega) \right] / \\
&&\!\!\!\!\!\!\!\!\!\!\!   \left( \left[\frac{\Gamma_m}{2}+\frac{\gamma(-\omega)}{2}+i (\delta+\lambda(-\omega)-\omega) \right] \! \left[\frac{\Gamma_m}{2}+\frac{\gamma(\omega)}{2}-i (\delta+\lambda(\omega)+\omega) \right] \right. \nonumber \\
&& \Bigg. - \Sigma(\omega) \Sigma^*(-\omega) \Bigg) . \nonumber 
\end{eqnarray}
Again, these expressions are quite involved. But nicely enough, the peaks look reasonably Lorentzian. So our procedure is to numerically find position, width and area of the red and blue sideband peaks, and compare these to  Lorentzian fits of experimental data.

%\newpage
%\include{to be deleted/equations holding cell}

\end{document}